\documentclass[twocolumn,showpacs,preprintnumbers,prd,superscriptaddress,nofootinbib]{revtex4-1}
\usepackage{graphicx}
\usepackage{epsf}
\usepackage{bm}
\usepackage{amsmath}
\usepackage{amsfonts}
\usepackage{amssymb}
\usepackage{epstopdf}
\usepackage{natbib}
\usepackage{hyperref}
\usepackage{color}
\usepackage{verbatim}
\usepackage{multirow}
\usepackage{bm}
\usepackage{hyperref}
\usepackage{float}
\usepackage{xcolor}

\definecolor{darkblue}{rgb}{0.0, 0.0, 0.55}
\definecolor{darkred}{rgb}{0.55, 0.0, 0.0}

\usepackage{hyperref}
\hypersetup{
    colorlinks=true, 
    linkcolor=darkblue,
    citecolor=darkblue,
    urlcolor=darkblue}
    
\makeatletter\let\expandableinput\@@input\makeatother

\begin{document}

\title{Modeling Uncertainties in Modified Gravity Predictions for the Stochastic Gravitational-Wave Background}

\author{Rodrigo Fraga}
\email{rodrigo.fraga@ufrgs.br}
\affiliation{Instituto de F\'{i}sica, Universidade Federal do Rio Grande do Sul, 91501-970 Porto Alegre RS, Brazil}

\author{Rafael C. Nunes}
\email{rafadcnunes@gmail.com}
\affiliation{Instituto de F\'{i}sica, Universidade Federal do Rio Grande do Sul, 91501-970 Porto Alegre RS, Brazil}
\affiliation{Divisão de Astrofísica, Instituto Nacional de Pesquisas Espaciais, Avenida dos Astronautas 1758, São José dos Campos, 12227-010, São Paulo, Brazil}

\begin{abstract}
We investigate the impact of modified gravity on the stochasticgravitational-wave background (SGWB) generated by a cosmological population of unresolved binary black hole mergers. We consider two complementary classes of beyond-General Relativity (GR) effects: waveform-generation modifications
described within the parametrized post-Einsteinian (ppE) framework and cosmological propagation effects associated with a modified gravitational-wave luminosity distance. Astrophysical uncertainties in the binary black hole population are consistently incorporated using a Power-Law plus Peak mass model
combined with a Madau--Dickinson merger-rate evolution. Using SGWB forecasts for Advanced LIGO at O5 design sensitivity, the Einstein Telescope (ET), Cosmic Explorer (CE), {and the combined CE+ET network}, we perform injection-recovery analyses jointly varying modified-gravity and astrophysical population parameters, including a joint injection in which waveform-generation and propagation deviations are present simultaneously. We show that frequency-dependent ppE corrections produce characteristic distortions in the SGWB spectral shape and can be meaningfully constrained by third-generation detectors, with the CE+ET network providing the tightest constraints. In contrast, modified propagation effects mainly induce smooth amplitude rescalings, are recovered with systematically lower accuracy, and exhibit stronger degeneracies with the merger-rate evolution. In the joint injection, the spectral-shape information is preserved while the amplitude-like parameters retain a residual mutual degeneracy. Our results demonstrate that future SGWB observations will provide a complementary probe of gravitational physics across cosmic history and may open new avenues for testing deviations from GR beyond individually resolved compact-binary events.

\end{abstract}

\maketitle

\section{Introduction}
\label{intro}

One century after the formulation of General Relativity (GR), the direct detection of gravitational waves (GWs) by the LIGO Scientific Collaboration and the Virgo Collaboration marked the beginning of a new era in observational physics \cite{LIGOScientific:2016aoc}. This observation has opened an independent window to probe the Universe, complementing traditional electromagnetic observations (see \cite{Bailes:2021tot,Cai:2017cbj} for a review). While most GW detections to date arise from individually resolved compact binary coalescences—such as binary black holes, binary neutron stars, and neutron star–black hole systems superposition of a large number of unresolved sources across cosmological distances is expected to generate a stochastic gravitational-wave background (SGWB) \cite{Allen:1996vm,Allen:1997ad,Regimbau:2011rp,Christensen:2018iqi,Renzini:2022alw}. This background is typically characterized by the dimensionless energy density spectrum, $\Omega_{\mathrm{GW}}(f)$, which encodes information about both the source population and the underlying cosmological model.

Despite significant observational efforts, the SGWB from the stellar binaries has not yet been directly detected, and current experiments have only placed upper bounds on its amplitude. The most recent constraints reported by the LIGO-Virgo-KAGRA (LVK) Collaboration set $\Omega_{\mathrm{GW}}(f) < 2.8 \times 10^{-9}$ at $f = 25\,\mathrm{Hz}$ (95\% C.L.), assuming a frequency-independent spectrum \cite{LIGOScientific:2025bgj}. The LVK Collaboration has also searched for an anisotropic gravitational-wave background \cite{LIGOScientific:2025bkz}.\footnote{Recent results from pulsar timing array experiments, in particular the NANOGrav Collaboration 15-year data set \cite{NANOGrav:2023gor,Agazie:2026tui,NANOGrav:2025gqp}, provide strong evidence for a common-spectrum stochastic process at nanohertz frequencies. While consistent with a SGWB, its origin is not yet established and may be attributed to a population of supermassive black hole binaries or alternative cosmological sources.}
However, the next generation of detectors -- including the Einstein Telescope (ET) \cite{ET:2025xjr}, Cosmic Explorer (CE) \cite{Evans:2021gyd}, and LISA \cite{LISA:2022yao} -- is expected to achieve the sensitivity required for a robust detection, potentially enabling precision measurements of the SGWB spectral shape across a wide range of frequencies.

In parallel, testing GR in new regimes remains a central goal of modern physics (see \cite{Berti:2015itd,Ishak:2018his,Clifton:2011jh} for a review). Although all GW observations so far are consistent with GR predictions, there are strong theoretical and observational motivations to consider deviations from GR, particularly on cosmological scales. Modifications to gravity can arise in attempts to explain the late-time acceleration of the Universe, address tensions within the standard cosmological model, or describe high-energy regimes in the early Universe. Such deviations generically affect GW propagation, leading to modifications in amplitude, phase, or polarization content over cosmological distances \cite{Nishizawa:2017nef,Belgacem:2018lbp,Nishizawa:2019rra}.
A comprehensive set of tests of GR based on the compact binary signals included in the GWTC-4.0 catalog was presented in \cite{LIGOScientific:2026qni,LIGOScientific:2026fcf,LIGOScientific:2026wpt}, with no statistically significant deviations from GR observed in this early stage of gravitational-wave tests of gravity. In this context, the stochastic background provides a particularly powerful and complementary probe. Unlike individual GW events, the SGWB encodes cumulative effects integrated over the entire cosmic history, rendering it especially sensitive to small deviations from GR that accumulate during propagation (see \cite{Nunes:2020rmr,Ezquiaga:2021ayr,Saffer:2020xsw} for representative studies at ground-based detector frequencies).

In this work, we adopt a parameterized post-Einsteinian (ppE) framework \cite{Yunes:2009ke,Cornish:2011ys,Tahura:2018zuq} to investigate generic deviations from GR in the SGWB. 
{The ppE approach parameterizes modifications to the generation of GWs beyond GR, particularly during the coalescence, merger, and ringdown phases of compact binary systems. In its original formulation, however, the ppE framework does not explicitly account for cosmological effects on GW propagation. In principle, such effects can also be incorporated within a similar phenomenological parameterization, allowing deviations from GR in the propagation of GWs to be consistently investigated.}
{It is important to note that some progress in the ppE framework has already been made in \cite{Saffer:2020xsw}. Here, we extend these efforts by developing a more general framework and, importantly, constructing a practical Bayesian analysis that accounts for realistic astrophysical uncertainties associated with the gravitational-wave sources.}
Rather than focusing on specific modified gravity theories, the ppE approach provides a model-independent description of possible deviations through phenomenological corrections to the gravitational-wave spectrum. Within this framework, we extend the standard calculation of the astrophysical SGWB by incorporating modifications that affect the gravitational waves, leading to corrections in the amplitude of $\Omega_{\mathrm{GW}}(f)$.

{In addition to source-generation effects, we consider a second and complementary class of deviations from GR, associated with the propagation of GWs over cosmological distances. These are parametrized through a modified gravitational-wave luminosity distance, controlled by a running Planck-mass parameter $\alpha_{M0}$~\cite{Nunes:2020rmr}, which reweights the contribution of distant sources to the background and thus acts on the SGWB. Because a given modified-gravity theory may simultaneously affect both the generation and propagation of GWs, we also perform a joint injection–recovery analysis with deviations in both sectors, assessing to what extent the SGWB can disentangle their respective effects.}

It is well known that uncertainties in the underlying compact-binary population have a significant impact on the predicted SGWB energy spectrum. In particular, assumptions regarding the black hole mass distribution, merger rate evolution, and redshift dependence can lead to substantial variations in both the amplitude and shape of $\Omega_{\mathrm{GW}}(f)$. In this work, we account for these astrophysical uncertainties following state-of-the-art prescriptions available in the literature (see \cite{Renzini2024popstock} and reference there). This allows us to assess the robustness of our results against variations in the source population.

We compute the SGWB spectrum by including contributions from compact binary coalescences, namely binary black holes, and evaluate the impact of modified gravity effects within the ppE framework. We then assess the detectability of these deviations using the sensitivity curves of current and future gravitational-wave detectors, including LIGO, ET, and CE. In particular, we analyze how ppE-induced modifications affect the amplitude and shape of $\Omega_{\mathrm{GW}}(f)$ and derive forecast constraints on the free parameters of the model considered in this work.

This paper is organized as follows. In Section \ref{model}, we introduce the theoretical framework for the SGWB in the presence of modified gravity. Section \ref{methodology} describes the methodology adopted in this work. Our main results and forecast constraints are presented in Section \ref{results}. Finally, Section \ref{final} summarizes our conclusions and discusses future perspectives.

\section{Gravitational-Wave Background from Compact Binary Coalescences in Modified Gravity}
\label{model}

The dimensionless energy density spectrum of the stochastic gravitational-wave background (SGWB) measured by a detector over a finite observation time can be written as \cite{Meacher:2015iua,Regimbau:2022mdu}
\begin{equation}
\Omega_{\rm GW}(f)
=
\frac{f^3}{T_{\rm obs}}
\frac{4\pi^2}{3H_0^2}
\sum_{i=1}^{N_{\rm ev}}
P_d(\Theta_i; f),
\label{eq:OmegaGW_discrete}
\end{equation}
where $f$ is the observed frequency, $T_{\rm obs}$ is the total observation time, $H_0$ is the Hubble constant today, and $N_{\rm ev}$ is the number of detected events during $T_{\rm obs}$. The quantity $P_d(\Theta_i; f)$ denotes the one-sided spectral power of the $i$-th event in the detector frame, with $\Theta_i$ representing the full set of source parameters.

In the frequency domain, the spectral power is defined as
\begin{equation}
P_d(\Theta_i; f)
=
\tilde h_+^2(\Theta_i; f)
+
\tilde h_\times^2(\Theta_i; f),
\label{eq:Pd_def}
\end{equation}
where $\tilde h_{+,\times}$ are the Fourier transforms of the two GW polarizations. This quantity has dimensions of time squared and encodes the spectral contribution of each individual source.

In the limit of long observation time and a large number of events, $T_{\rm obs} \to \infty$ and $N_{\rm ev} \to \infty$, the measured spectrum converges to the ensemble average $\bar{\Omega}_{\rm GW}$. In this regime, the discrete sum can be replaced by a statistical average over the source population,
\begin{equation}
\frac{1}{T_{\rm obs}}
\sum_{i=1}^{N_{\rm ev}}
P_d(\Theta_i; f)
\;\longrightarrow\;
R
\int d\Theta \,
p_d(\Theta|\Lambda)\,
P_d(\Theta; f),
\label{eq:sum_to_integral}
\end{equation}
where $R \equiv dN/dt$ is the total event rate in the detector frame, $p_d(\Theta|\Lambda)$ is the normalized probability distribution of source parameters, and $\Lambda$ denotes the set of population hyperparameters.

Substituting Eq.~\eqref{eq:sum_to_integral} into Eq.~\eqref{eq:OmegaGW_discrete}, we obtain
\begin{equation}
\bar{\Omega}_{\rm GW}(\Lambda; f)
=
f^3
\frac{4\pi^2}{3H_0^2}
R
\int d\Theta \,
p_d(\Theta|\Lambda)\,
P_d(\Theta; f).
\label{eq:OmegaGW_mean}
\end{equation}

It is convenient to separate the redshift dependence by writing $\Theta = (z,\theta)$, where $\theta$ denotes intrinsic source parameters (e.g., masses, spins, orientations). Assuming statistical independence between $z$ and $\theta$, we factorize the distribution as
\begin{equation}
p_d(\Theta|\Lambda) = p(z)\,p(\theta|z,\Lambda),
\qquad
\int dz\,p(z)=1.
\end{equation}
The rate density can then be written as $R(z) = R\,p(z)$, allowing Eq.~\eqref{eq:OmegaGW_mean} to be expressed as
\begin{equation}
\bar{\Omega}_{\rm GW}(f)
=
f^3
\frac{4\pi^2}{3H_0^2}
\int dz\,
R(z)
\,
\big\langle P_d(z; f) \big\rangle,
\label{eq:OmegaGW_shells}
\end{equation}
where the average over intrinsic parameters is defined as
\begin{equation}
\big\langle P_d(z; f) \big\rangle
=
\int d\theta \,
p(\theta|z,\Lambda)\,
P_d(\theta,z; f).
\label{eq:Pd_average}
\end{equation}

More generally, for any quantity ${\cal Q}(\theta)$, the population average is
\begin{equation}
\langle {\cal Q} \rangle
=
\int d\theta \, p(\theta)\, {\cal Q}(\theta),
\qquad
\int d\theta\,p(\theta)=1.
\end{equation}
A relevant example is the source-frame energy spectrum $dE_s/df_s$, whose population average reads
\begin{equation}
\left\langle \frac{dE_s}{df_s} \right\rangle
=
\int d\theta \, p(\theta)\,
\frac{dE_s}{df_s}(\theta).
\end{equation}
The source-frame frequency is related to the observed one via $f_s = (1+z)f$.

\vspace{0.2cm}

\noindent
\textit{ppE modifications.—}
Within the ppE framework, deviations from GR are introduced at the waveform level. In the frequency domain, the waveform can be written as
\begin{equation}
\tilde{h}_{A}(f)
=
\tilde{h}^{\rm GR}_{A}(f)\,
\left(1+\alpha_{ppE}\,u^a\right)
e^{i\beta u^b},
\qquad A=+,\times,
\label{eq:ppE_waveform}
\end{equation}
where
\begin{equation}
u = \left(\pi \mathcal{M} f_s\right)^{1/3},
\qquad
f_s = (1+z)f,
\end{equation}
and $\mathcal{M}$ is the chirp mass.

Since $\Omega_{\rm GW}$ depends on $|\tilde h|^2$, phase corrections do not contribute at leading order, i.e.,
\begin{equation}
\left| e^{i\beta u^b} \right|^2 = 1,
\end{equation}
and only amplitude modifications affect the SGWB spectrum. The spectral power becomes
\begin{equation}
P_d^{\rm ppE}(\Theta;f)
=
P_d^{\rm GR}(\Theta;f)\,
\left(1+\alpha_{ppE} u^a\right)^2.
\label{eq:Pd_ppE}
\end{equation}

In the perturbative regime, $|\alpha_{ppE} u^a|\ll1$, this reduces to
\begin{equation}
P_d^{\rm ppE}
\simeq
P_d^{\rm GR}
\left(1 + 2\alpha_{ppE} u^a \right).
\end{equation}

Substituting into Eq.~\eqref{eq:OmegaGW_mean}, we obtain
\begin{align}
\bar\Omega_{\rm GW}^{\rm ppE}(f)
&=
f^3
\frac{4\pi^2}{3H_0^2}
\int dz\,R(z)
\int d\theta\,
\nonumber\\
&\quad \times
p(\theta|z,\Lambda)\,
P_d^{\rm GR}(z,\theta;f)
\left(1+\alpha_{ppE} u^a\right)^2.
\end{align}

Expanding to leading order in $\alpha_{ppE}$, we find
\begin{align}
\bar\Omega_{\rm GW}^{\rm ppE}(f)
&\simeq
\bar\Omega_{\rm GW}^{\rm GR}(f)
+
2\alpha_{ppE}
f^3
\frac{4\pi^2}{3H_0^2}
\int dz\,R(z)
\int d\theta\,
\nonumber\\
&\quad \times
p(\theta|z,\Lambda)\,
P_d^{\rm GR}(z,\theta;f)\,
u^a.
\label{eq:OmegaGW_ppE_linear}
\end{align}

Using
\begin{equation}
u =
(\pi \mathcal{M})^{1/3}
(1+z)^{1/3}
f^{1/3},
\end{equation}
the final result can be written in compact form as
\begin{align}
\bar\Omega_{\rm GW}^{\rm ppE}(f)
=&
\bar\Omega_{\rm GW}^{\rm GR}(f)
\left[ \right.
1
\nonumber\\
& \left.+
2\alpha_{ppE}
f^{a/3}
\left\langle
(\pi \mathcal{M})^{a/3}
(1+z)^{a/3}
\right\rangle_E
\right],
\label{eq:OmegaGW_ppE_final}
\end{align}
where $\langle \cdots \rangle_E$ denotes an energy-weighted average over the source population,
\begin{equation}
\label{source_population}
\langle X \rangle_E
=
\frac{
\int dz\,R(z)
\int d\theta\,p(\theta|z)\,
P_d^{\rm GR}(z,\theta;f)\,X
}{
\int dz\,R(z)
\int d\theta\,p(\theta|z)\,
P_d^{\rm GR}(z,\theta;f)
}.
\end{equation}

Therefore, at leading order, only amplitude corrections contribute to $\Omega_{\rm GW}$, inducing a characteristic frequency-dependent rescaling governed by the ppE parameters. This result highlights the SGWB as a sensitive probe of beyond-GR effects integrated over cosmic history, opening the possibility of constraining modified gravity scenarios.

\begin{figure*}
\centering
\includegraphics[width=0.45\textwidth]{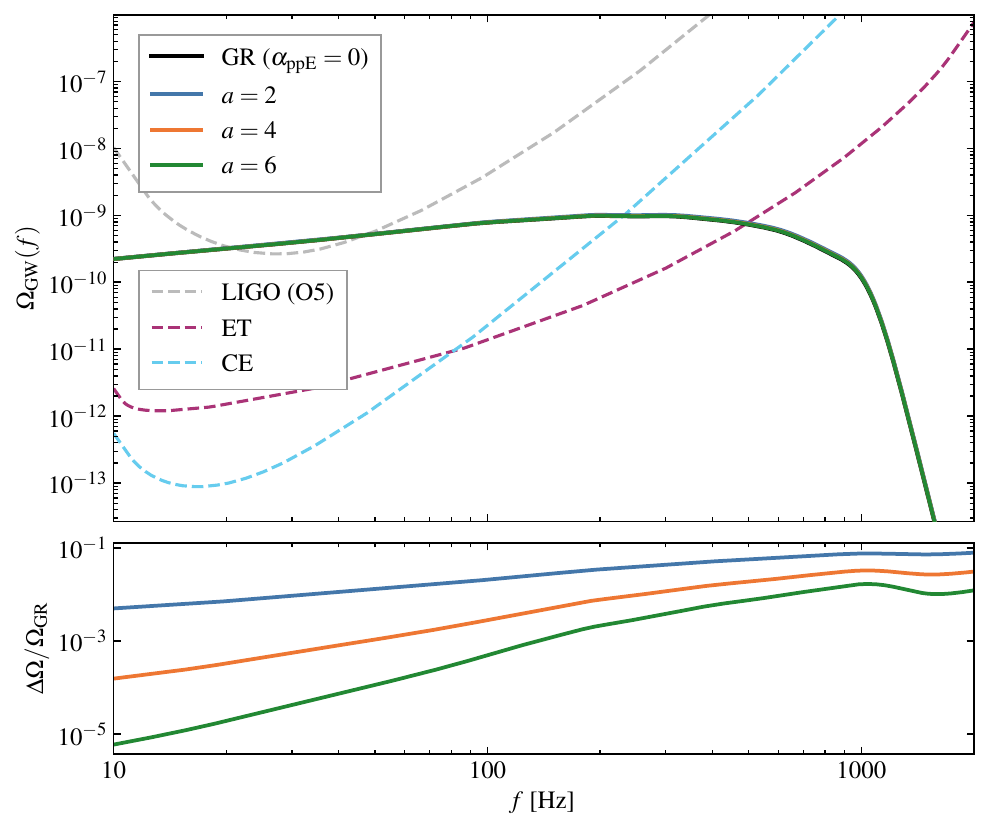}
\includegraphics[width=0.45\textwidth]{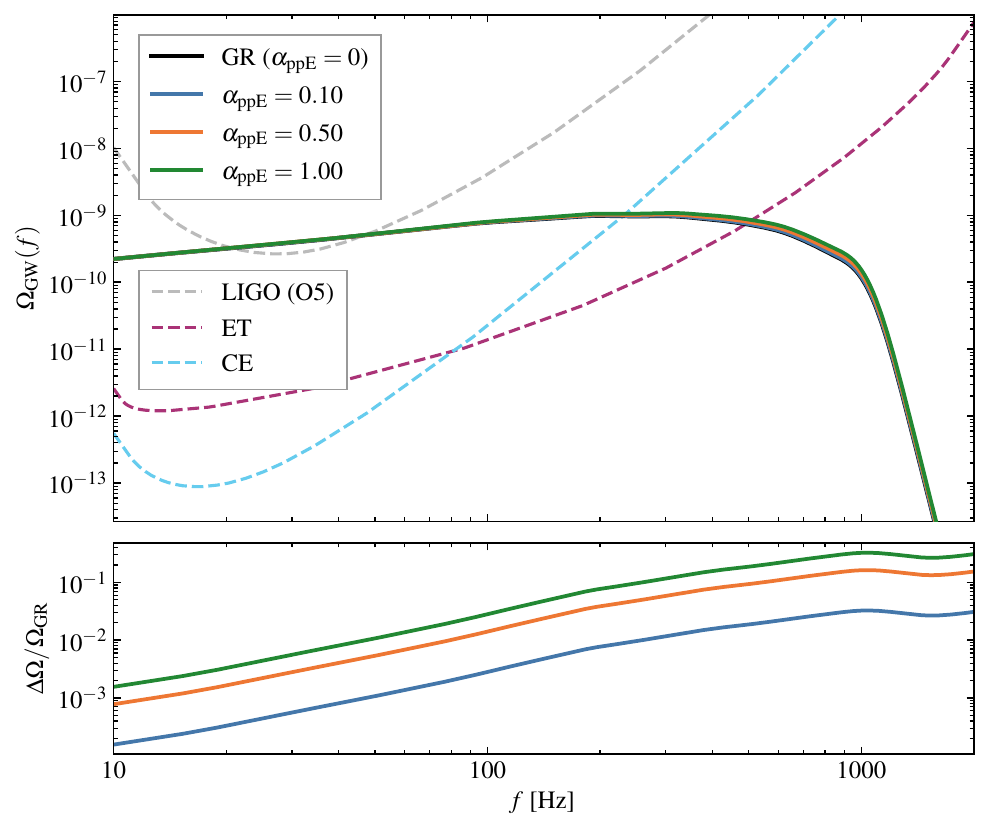}
\caption{Impact of ppE amplitude corrections on the SGWB energy-density
spectrum, $\Omega_{\rm GW}(f)$. \textit{Left panels:} the amplitude parameter is fixed to $\alpha_{\rm ppE} = 0.10$, while different PN orders,
$a \in [2,4,6]$, are considered and compared with the GR prediction
($\alpha_{\rm ppE} = 0$). \textit{Right panels:} the PN order is fixed to
$a = 4$, while the amplitude parameter varies as $\alpha_{\rm ppE} \in [0.10, 0.50, 1.00]$. In the upper panels, the dashed lines show the projected sensitivities of Advanced LIGO (O5), ET and CE to a stochastic background, obtained from the detector noise power spectral
densities distributed with the \texttt{bilby} package~\cite{Ashton_2019}. The lower panels show the corresponding relative deviation,
$\Delta\Omega/\Omega_{\rm GR} \equiv
(\Omega^{\rm ppE}_{\rm GW} - \Omega^{\rm GR}_{\rm GW})/\Omega^{\rm GR}_{\rm GW}$,
for each scenario.}
\label{plots_ppe}
\end{figure*}

Figure \ref{plots_ppe} shows the impact of ppE amplitude corrections on the SGWB energy-density spectrum for different values of the Post-Newtonian (PN) exponent $a$ and amplitude parameter $\alpha_{\rm ppE}$. Negative PN orders ($a<0$) produce larger deviations at low frequencies, while positive PN orders ($a>0$) mainly affect the high-frequency regime. In particular, the case $a=-8$ generates a strong enhancement at low frequencies, whereas $a=4$ leads to deviations that grow with frequency.

The right panels show that increasing $\alpha_{\rm ppE}$ amplifies the deviation approximately linearly, without significantly altering the overall spectral shape. Overall, the ppE corrections act as a frequency-dependent rescaling of the GR spectrum, demonstrating the sensitivity of the SGWB to modified-gravity effects accumulated over the cosmic population of compact binary mergers. {In both panels of Figure \ref{plots_ppe}, we also show the sensitivity curves of Advanced LIGO (O5), ET, and CE to a stochastic gravitational-wave background, derived from injection simulations.}

\subsection{Modified gravitational-wave propagation (MGWP)}
\label{cosmo}

In addition to the modifications arising from the strong-gravity regime during the inspiral phase, as discussed previously, we also consider deviations from GR that affect the propagation of GWs. Such effects can be parameterized through a modified GW--EM luminosity-distance relation \cite{Nishizawa:2017nef,Mancarella:2021ecn,LIGOScientific:2025jau,LIGOScientific:2026uyd}. These constraints probe the hypothesis that gravity may depart from the predictions of GR on cosmological scales, which can be interpreted as evidence for an effective dark energy component or modified gravitational dynamics at late times.

For a given event observed at redshift $z_i$, the (unpolarized) spectral power measured in the detector frame can be expressed as
\begin{equation}
P_d(\Theta_i; f)
\propto
\frac{\mathcal{A}_{\rm src}^2(f_s;\theta_i)}{\left[d_L^{\rm GW}(z_i)\right]^2},
\label{eq:Pd_dLGW_theta}
\end{equation}
where $\mathcal{A}_{\rm src}(f_s;\theta_i)$ denotes the intrinsic source amplitude evaluated at the source-frame frequency $f_s = (1+z_i)f$, and
\begin{equation}
\Theta_i \equiv (z_i, \theta_i)
\end{equation}
represents the full set of source parameters, with $\theta_i$ encoding intrinsic properties such as masses, spins, and orbital orientation.

In theories beyond GR in which modifications affect only the propagation of gravitational waves, the waveform at emission remains unchanged, while deviations arise through the distance scaling of the amplitude. In this case, the standard electromagnetic luminosity distance $d_L^{\rm EM}(z)$ is replaced by an effective gravitational-wave luminosity distance $d_L^{\rm GW}(z)$, leading to a modified amplitude decay.

As a consequence, the SGWB accumulated over an observation time $T_{\rm obs}$ acquires an explicit dependence on $d_L^{\rm GW}(z)$,
\begin{equation}
\Omega_{\rm GW}(f)
\propto
\sum_{i=1}^{N_{\rm ev}}
\frac{\mathcal{A}_{\rm src}^2(f_s;\theta_i)}{\left[d_L^{\rm GW}(z_i)\right]^2}.
\label{eq:OmegaGW_dLGW_sum_theta}
\end{equation}

This modification alters the overall normalization of the SGWB spectrum, since contributions from sources at different redshifts are reweighted relative to the General Relativity prediction. In particular, as the difference between $d_L^{\rm GW}(z)$ and $d_L^{\rm EM}(z)$ accumulates over cosmic time, distant sources can be either enhanced or suppressed depending on the underlying theory of gravity.

At the same time, the redshift distribution of events is governed by the astrophysical merger rate and its evolution, introducing an intrinsic dependence on population properties. Therefore, modified GW propagation generically leads to degeneracies between cosmological effects and astrophysical uncertainties, such as the merger rate evolution, mass function, and redshift-dependent population synthesis.

These effects can be incorporated in a compact form by expressing the modified spectral power as a rescaling of the GR prediction,
\begin{equation}
P_d(\Theta; f)
=
P_d^{\rm GR}(\Theta; f)\,
\left[
\frac{d_L^{\rm EM}(z)}{d_L^{\rm GW}(z)}
\right]^2,
\label{eq:Pd_rescaling_theta}
\end{equation}
where $P_d^{\rm GR}$ denotes the spectral power computed assuming standard propagation.

In the limit of a large number of unresolved events, the SGWB spectrum can then be written as
\begin{equation}
\Omega_{\rm GW}(f)
=
\Omega_{\rm GW}^{\rm GR}(f)\,
\left\langle
\left[
\frac{d_L^{\rm EM}(z)}{d_L^{\rm GW}(z)}
\right]^2
\right\rangle_E,
\label{eq:OmegaGW_effective_theta}
\end{equation}
where $\Omega_{\rm GW}^{\rm GR}(f)$ is the standard prediction and $\langle \cdots \rangle_E$ denotes an energy-weighted average over the source population, eq. (\ref{source_population}).

{In particular, the relevant correction factor associated with modified propagation is given by
\begin{align}
&\left\langle
\left[
\frac{d_L^{\rm EM}(z)}{d_L^{\rm GW}(z)}
\right]^2
\right\rangle_E =
\nonumber\\
&\frac{
\int dz \, d\theta \;
R(z)\,p(\theta|z)\,
P_d^{\rm GR}(z,\theta; f)\,
\left[
\frac{d_L^{\rm EM}(z)}{d_L^{\rm GW}(z)}
\right]^2
}{
\int dz \, d\theta \;
R(z)\,p(\theta|z)\,
P_d^{\rm GR}(z,\theta; f)
}.
\label{eq:dL_average_theta}
\end{align}
}
This formulation makes explicit that deviations from GR in the propagation sector enter the SGWB as a redshift-dependent reweighting of the source population. As a result, the SGWB provides a sensitive probe of cumulative propagation effects across cosmic history, while simultaneously requiring careful treatment of astrophysical uncertainties in order to disentangle modified gravity signatures from population-driven variations.

\begin{figure}
\centering
\includegraphics[width=0.45\textwidth]{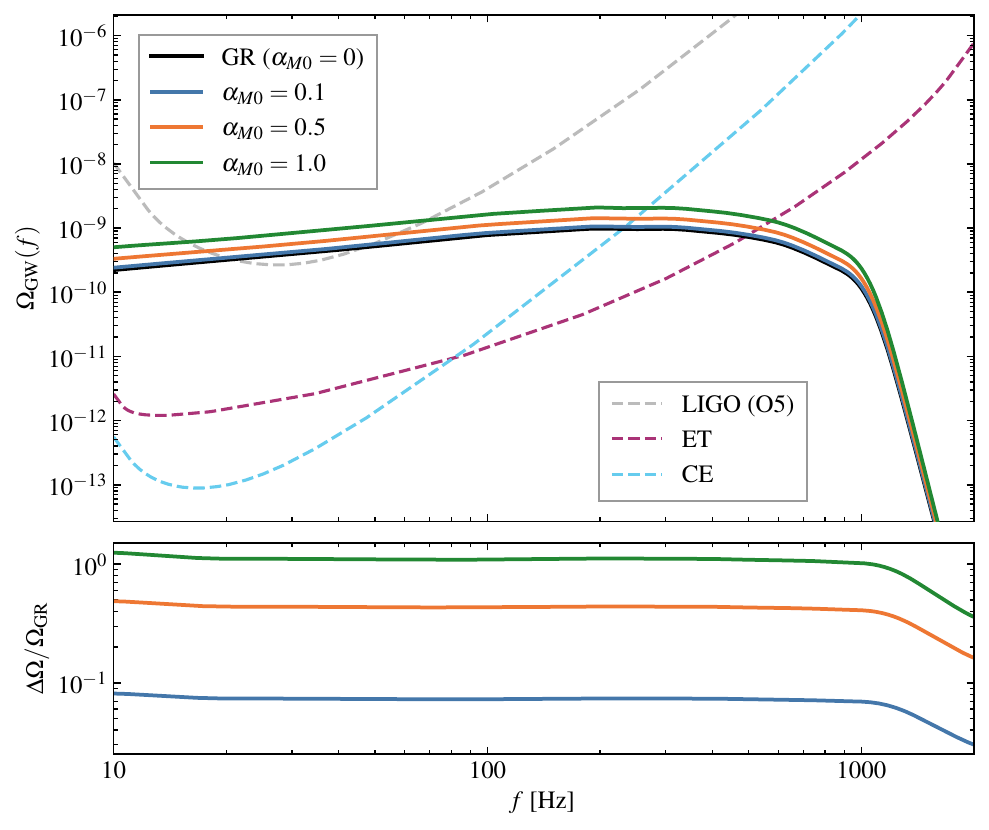}
\caption{SGWB energy-density spectrum, $\Omega_{\rm GW}(f)$, for different
values of the running Planck mass parameter $\alpha_{M0}$. The black curve
corresponds to the standard GR prediction ($\alpha_{M0} = 0$), while the
colored curves show the cases $\alpha_{M0} = 0.1$, $0.5$, and $1.0$. As in
Fig.~\ref{plots_ppe}, the dashed lines indicate the projected
sensitivities of Advanced LIGO (O5), ET and CE, derived from the noise power spectral densities available in the \texttt{bilby} package~\cite{Ashton_2019}. The lower panel displays the relative deviation, $|\Delta\Omega|/\Omega_{\rm GR}$, with respect to GR. Cosmological parameters are fixed to $\Omega_m = 0.31$,
$\Omega_r = 9.2 \times 10^{-5}$, and $\Omega_\Lambda = 1 - \Omega_m$.}
\label{plots_dl_dg}
\end{figure}

As discussed in \cite{Belgacem:2018lbp}, the ratio between the gravitational-wave and electromagnetic luminosity distances can be interpreted in terms of the evolution of an effective gravitational coupling strength,
\begin{equation}
\frac{d_L^{\rm EM}(z)}{d_L^{\rm GW}(z)}
= \sqrt{\frac{G_{\rm eff}(z)}{G_{\rm eff}(z=0)}} .
\end{equation}

Equivalently, this ratio can be expressed as
\begin{equation}
\frac{d_L^{\rm EM}(z)}{d_L^{\rm GW}(z)} = 
\exp\left[
\int_0^z \frac{dz'}{1+z'} \alpha_M(z')
\right],
\label{eq:Geff_alphaM}
\end{equation}
where $\alpha_M(z)$ denotes the running Planck mass parameter \cite{Lagos:2019kds}. This quantity characterizes the time evolution of the effective Planck mass and, consequently, encodes possible deviations from GR in the propagation of gravitational waves over cosmological distances.

A commonly adopted parametrization is given by \cite{Alonso:2016suf,Bellini:2015xja}
\begin{equation}
\alpha_M(z) =
\alpha_{M0}
\frac{\Omega_{\rm DE}(z)}{\Omega_{\rm DE}(0)},
\end{equation}
where $\Omega_{\rm DE}(z)$ denotes the dark energy density fraction as a function of redshift. This parametrization ensures that deviations from GR become relevant only at late times (low redshifts), when dark energy starts to dominate the cosmic expansion. In the limit $\alpha_{M0}=0$, the standard GR prediction is fully recovered, implying identical gravitational-wave and electromagnetic luminosity distances.

Figure \ref{plots_dl_dg} shows the SGWB energy-density spectrum for different values of the running Planck mass parameter $\alpha_{M0}$. In contrast to the ppE corrections discussed previously, the effect of modified propagation mainly produces an overall rescaling of the SGWB amplitude, while preserving the general spectral shape predicted by GR.

As expected from Eq.~\eqref{eq:OmegaGW_effective_theta}, positive values of $\alpha_{M0}$ enhance the SGWB amplitude relative to the GR prediction. This behavior reflects the modified GW luminosity distance relation, which effectively reduces the damping of gravitational waves during propagation and increases the contribution from distant sources. Consequently, the cumulative SGWB signal becomes progressively larger as $\alpha_{M0}$ increases.

The bottom panel shows the relative deviation $|\Delta\Omega|/\Omega_{\rm GR}$, which remains approximately constant over most of the frequency range. This nearly scale-independent behavior indicates that modified GW propagation primarily acts as a cosmological reweighting of the source population, rather than introducing a strong intrinsic frequency dependence. Small departures from a perfectly constant ratio at high frequencies arise from the interplay between the propagation effect and the redshift distribution of the contributing compact binary population.

\section{Methodology}
\label{methodology}

In this section, we briefly describe the methodology that will be used to forecast the free parameters of the theoretical framework outlined above, as well as the underlying astrophysical modeling adopted to estimate the stochastic gravitational-wave background spectrum, $\Omega_{\rm GW}(f)$.

The binary black hole (BBH) population adopted in this work is modeled through a phenomenological framework that combines the Power-Law plus Peak (PLPP) prescription for the mass distribution with the Madau--Dickinson (MD) parameterization for the redshift evolution of the merger rate. This approach is implemented through the \texttt{gwpopulation} package using the \texttt{SinglePeakSmoothedMassDistribution} model~\cite{Talbot2019}, and is broadly consistent with the population constraints inferred from the GWTC-3 catalog~\cite{Abbott2023pop}.

For the primary black hole mass, $m_1$, the probability density is described by a mixture model consisting of a truncated power-law component and an additional Gaussian peak,
\begin{align}
p(m_1 \mid \Lambda_m) =& \left[(1 - \lambda_{peak})\mathcal{P}(m_1 \mid \alpha_{\rm{IMF}}, m_{max}) \right.\nonumber\\
&+ \lambda_{peak} \left.\mathcal{G}(m_1 \mid \mu_{peak}, \sigma_{peak})\right]\nonumber\\
& \times S(m_1 \mid m_{min}, \delta_m),
\label{eq:plpp_mass}
\end{align}
where $\alpha_{\rm{IMF}}$ determines the slope of the power-law component, while $\mu_{peak}$ and $\sigma_{peak}$ characterize the mean and width of the Gaussian contribution. The parameter $\lambda_{peak}$ specifies the fraction of systems belonging to this excess high-mass component, and the smoothing function $S(m_1 \mid m_{min}, \delta_m)$ imposes a gradual suppression below the minimum mass threshold $m_{min}$ over a characteristic width $\delta_m$. The upper truncation is controlled by $m_{max}$.

The secondary mass is parameterized through the mass ratio $q = m_2/m_1 \leq 1$, which is assumed to follow a power-law distribution with spectral index $\beta$. Together, these parameters define a flexible model capable of capturing both the broad astrophysical mass hierarchy of BBHs and possible excesses associated with specific formation channels, such as pulsational pair-instability remnants.

Since the gravitational-wave energy emitted by each binary scales strongly with the chirp mass, the abundance of massive systems plays a central role in determining the normalization and spectral shape of the SGWB. In particular, the Gaussian peak can significantly enhance the SGWB amplitude and alter the location of its high-frequency turnover.

The redshift evolution of the BBH merger rate is modeled through the Madau--Dickinson functional form~\cite{Madau2014},
\begin{equation}
R(z \mid R_0, \gamma, \kappa, z_{peak}) =
R_0 \frac{(1 + z)^{\gamma}}
{1 + \left(\dfrac{1 + z}{1 + z_{peak}}\right)^{\kappa}},
\label{eq:madau_dickinson}
\end{equation}
where $R_0$ denotes the local merger rate density at $z=0$, $\gamma$ controls the low-redshift rise, $\kappa$ governs the suppression at high redshift, and $z_{peak}$ specifies the redshift at which the merger rate reaches its maximum.

The corresponding redshift probability density is given by
\begin{equation}
p(z) \propto \frac{1}{1+z}\frac{dV_c}{dz}R(z),
\label{eq:pz}
\end{equation}
where $dV_c/dz$ is the differential comoving volume element, computed assuming the Planck 2015 cosmological parameters~\cite{Planck2015}. This formulation naturally incorporates both cosmological expansion and source-frame time dilation effects.

The full set of BBH population hyperparameters considered in this analysis is therefore
\begin{align}
\Lambda = \{&\alpha_{\rm{IMF}}, \beta, m_{min}, m_{max}, \delta_m, \lambda_{peak}, \mu_{peak}, \sigma_{peak}, \nonumber\\
& R_0, \gamma, \kappa, z_{peak}\}.
\end{align}

These parameters jointly determine the astrophysical prior for the SGWB signal, linking compact-object formation, binary evolution, and cosmic star-formation history to the observed stochastic background.

For an unresolved cosmological BBH population, the ensemble-averaged SGWB is expressed as~\cite{Phinney:2001,Meacher2015}
\begin{equation}
\bar{\Omega}_{GW}(\Lambda; f) =
\frac{4\pi^2 f^3}{3H_0^2}
R \int d\Theta\, p_d(\Theta \mid \Lambda)\, P_d(\Theta; f),
\label{eq:omega_ensemble}
\end{equation}
where $\Theta$ represents the full set of binary parameters in the detector frame, $p_d(\Theta \mid \Lambda)$ is the population probability density, and
\begin{equation}
P_d(\Theta; f)=|\tilde{h}_+|^2+|\tilde{h}_\times|^2
\end{equation}
is the detector-frame strain power spectrum.

Directly reevaluating this multidimensional integral for each new astrophysical hyperparameter configuration would be computationally prohibitive. To overcome this limitation, we employ the \textsc{PopStock} framework~\cite{Renzini2024popstock}, which uses an importance-sampling reweighting strategy:
\begin{equation}
\int d\Theta\, p_d(\Theta\mid\Lambda_i)P_d(\Theta)
\approx
\sum_j w_i(\Theta_j) P_d(\Theta_j),
\label{eq:importance_sampling}
\end{equation}
with statistical weights
\begin{equation}
w_i(\Theta_j)=
\frac{p_d(\Theta_j\mid\Lambda_i)}
{p_d(\Theta_j\mid\Lambda_0)},
\label{eq:importance_sampling2}
\end{equation}
where $\Lambda_0$ denotes a fiducial reference population and $\{\Theta_j\}$ is a precomputed Monte Carlo sample generated from it.

This methodology allows all waveform-dependent quantities to be calculated only once, while variations in the BBH population are efficiently encoded through the reweighting factors. As a result, SGWB predictions for different population models can be produced with dramatically reduced computational cost, making large-scale parameter forecasting feasible. The reliability of this procedure is quantified through the effective sample size $N_{\rm eff}$, which monitors the statistical robustness of the reweighted sample~\cite{Renzini2024popstock,Giarda2025}.

The detector-frame strain power is computed using the frequency-domain inspiral-merger-ringdown approximant \texttt{IMRPhenomD}~\cite{Khan2016}, implemented through the \textsc{bilby}/\textsc{LALSuite} infrastructure \cite{Ashton_2019}. This waveform model has become standard in SGWB population analyses~\cite{Renzini2024popstock,Giarda2025}, and uncertainties associated with alternative waveform prescriptions remain subdominant compared to current astrophysical uncertainties.

In this analysis, BBHs are assumed to be non-spinning, which provides a computationally efficient baseline while preserving the dominant features relevant for SGWB forecasts.

The SGWB spectrum is evaluated on a logarithmically spaced detector-frame frequency grid spanning $f \in [10,2048]\,{\rm Hz}$,
consistent with the projected sensitivity range of third-generation gravitational-wave observatories such as Cosmic Explorer and Einstein Telescope~\cite{Reitze2019,ET:2025xjr}. A reference frequency of $f_{ref}=25\,{\rm Hz}$ is adopted.

The detector configurations adopted in this analysis are Advanced LIGO at O5 design sensitivity~\cite{LIGOScientific:2025bgj}, the Einstein Telescope (ET)~\cite{ET:2025xjr}, Cosmic Explorer (CE)~\cite{Evans:2021gyd,Reitze2019}, {and the combined CE+ET network. Their sensitivity to a stochastic background is shown by the dashed lines in Figs.~\ref{plots_ppe} and~\ref{plots_dl_dg}, computed from the detector noise power spectral densities distributed with the \texttt{bilby} package~\cite{Ashton_2019} and expressed in terms of $\Omega_{\rm GW}(f)$ over the band relevant for this work, $[10\,\mathrm{Hz} - 1\,\mathrm{kHz}]$.} Third-generation detectors improve the accessible SGWB energy density by roughly three orders of magnitude with respect to Advanced LIGO in the decade below $\sim 100$~Hz, precisely where the BBH contribution to $\Omega_{\rm GW}$ is strongest and where, as discussed above, most of the constraining power of the analysis is concentrated. {For the CE+ET configuration the sensitivity is dominated by the cross-correlation between two widely separated sites; the corresponding overlap reduction function decorrelates above a few tens of hertz~\cite{Allen:1997ad,Christensen:2018iqi}, so the combined curve does not lie below the individual CE curve over the whole band.}

Although the full SGWB spectrum is generated across this broad interval, the majority of the constraining power is concentrated below approximately $200\,{\rm Hz}$. In this lower-frequency regime, the stochastic BBH signal is strongest, detector sensitivity is optimal, and statistical fluctuations due to finite source counts are minimized. At higher frequencies, the merger-ringdown suppression of BBH spectra, rising detector noise, and increasing Poisson variance progressively reduce the available information content~\cite{Giarda2025}.

Consequently, the low-frequency sector dominates the inference of BBH population hyperparameters and serves as the primary driver of the forecasting analysis performed in this work.

To assess the sensitivity of next-generation detectors to departures from general relativity, we simulate two distinct SGWB signals. The first, hereafter referred to as the GR signal, is generated within standard general relativity, with no ppE correction applied, corresponding to $\alpha_{\rm ppE}=0$. The second, referred to as the ppE signal, includes a small but non-negligible deviation from GR, characterized by $\alpha_{\rm ppE}=0.5$. We then repeat the same analysis within the modified GW propagation framework, adopting $\alpha_{M0}=0.5$.

Both signals are generated using a modified implementation of the \texttt{PyGWB} package~\cite{Renzini2023pygwb}, adapted to evaluate the ppE-corrected spectrum given by Eq.~\eqref{eq:OmegaGW_ppE_final} instead of the standard GR expression. Each simulated signal is subsequently subjected to a parameter-estimation analysis, in which the ppE parameters $a$, $\alpha_{\rm ppE}$, and $\alpha_{M0}$ are allowed to vary jointly with the BBH population hyperparameters $\Lambda$.

In this work, we focus exclusively on BBH population as the dominant astrophysical contribution in the frequency range of interest. Although binary neutron star (BNS) and neutron star--black hole (NSBH) systems also contribute to the SGWB, their predicted amplitudes are significantly smaller than the BBH component.
This behavior arises from the stronger GW emission produced by massive BBH systems, whose larger chirp masses dominate the SGWB amplitude over the frequency range relevant for this work. In particular, BBHs provide the leading contribution below a few hundred hertz, where detector sensitivity and statistical constraining power are maximal. Including BNS and NSBH populations would considerably enlarge the astrophysical parameter space without qualitatively modifying the beyond-GR signatures investigated here. Therefore, we restrict the analysis to a cosmological population of unresolved BBH mergers, which captures the dominant contribution to the SGWB while providing a computationally efficient baseline for forecasting modified-gravity effects.

\section{Results}
\label{results}

\begin{table*}[ht]
\centering
\caption{Summary of the BBH population hyperparameters and modified gravity parameters used in this analysis, together with their fiducial (injected) values and the posteriors recovered for each detector configuration, quoted as median values with $1\sigma$ credible intervals. The population model combines the Power-Law plus Peak (PLPP) mass distribution with the Madau--Dickinson (MD) merger-rate parametrization, as described in Sec.~\ref{methodology}. Results are shown for four injections: a pure GR signal ($\alpha_{\rm ppE}=0$); a ppE waveform-generation signal with $\alpha_{\rm ppE}=0.5$ and $a=4$; a modified GW propagation (MGWP) signal with $\alpha_{M0}=0.5$; and a joint ppE+MGWP signal in which both classes of deviation are injected simultaneously, with $(a,\alpha_{\rm ppE},\alpha_{M0})=(4,0.5,0.5)$. Dashes indicate parameters that are not present in a given injection.}
\label{tab:inf_geral}
\renewcommand{\arraystretch}{1.5}
\setlength{\tabcolsep}{3pt}
\begin{tabular}{lccccccc}
\hline
Detector & $a$ & $\alpha_\mathrm{ppE}$ & $\alpha_\mathrm{M0}$ & $\alpha_\mathrm{IMF}$ & $\lambda_\mathrm{peak}$ & $\mathcal{R}\ [\mathrm{Gpc}^{-3}\mathrm{yr}^{-1}]$ & $\gamma$ \\
\hline
\multicolumn{8}{c}{\bf GR Signal Injection} \\
\hline
Injected Values & $4$ & $0$ & -- & $3.5$ & $0.03$ & $17$ & $3$ \\
\hline
CE+ET & $3.569_{-1.071}^{+0.965}$ & $0.03648_{-0.02759}^{+0.03160}$ & -- & $3.278_{-0.144}^{+0.146}$ & $0.03372_{-0.00349}^{+0.00333}$ & $19.33_{-1.23}^{+1.29}$ & $2.754_{-0.12}^{+0.115}$ \\
CE & $3.417_{-1.075}^{+1.124}$ & $0.04044_{-0.03022}^{+0.03183}$ & -- & $3.255_{-0.197}^{+0.194}$ & $0.03490_{-0.00493}^{+0.00487}$ & $19.73_{-1.66}^{+1.72}$ & $2.713_{-0.154}^{+0.150}$ \\
ET & $2.890_{-1.435}^{+1.440}$ & $0.04568_{-0.03170}^{+0.03237}$ & -- & $4.043_{-1.356}^{+1.571}$ & $0.06573_{-0.01985}^{+0.02167}$ & $26.29_{-7.23}^{+7.34}$ & $2.014_{-0.604}^{+0.613}$ \\
LIGO & $2.549_{-1.651}^{+1.612}$ & $0.04904_{-0.03281}^{+0.03266}$ & -- & $3.945_{-2.432}^{+2.697}$ & $0.04889_{-0.03164}^{+0.03130}$ & $19.88_{-9.09}^{+8.60}$ & $2.083_{-0.829}^{+0.853}$ \\
\hline
\multicolumn{8}{c}{\bf ppE Signal Injection} \\
\hline
Injected Values & $4$ & $0.5$ & -- & $3.5$ & $0.03$ & $17$ & $3$ \\
\hline
CE+ET & $4.005_{-0.621}^{+0.647}$ & $0.5335_{-0.2052}^{+0.2172}$ & -- & $3.277_{-0.141}^{+0.143}$ & $0.03196_{-0.00573}^{+0.00595}$ & $19.22_{-1.19}^{+1.18}$ & $2.772_{-0.113}^{+0.116}$ \\
CE & $3.955_{-0.699}^{+0.742}$ & $0.5149_{-0.2309}^{+0.2296}$ & -- & $3.313_{-0.209}^{+0.202}$ & $0.03328_{-0.00654}^{+0.00680}$ & $19.12_{-1.73}^{+1.72}$ & $2.779_{-0.166}^{+0.174}$ \\
ET & $4.063_{-0.704}^{+0.680}$ & $0.6856_{-0.3859}^{+0.4064}$ & -- & $4.163_{-1.374}^{+1.533}$ & $0.06584_{-0.02193}^{+0.02289}$ & $26.19_{-6.96}^{+7.49}$ & $2.033_{-0.603}^{+0.589}$ \\
LIGO & $2.955_{-1.509}^{+1.365}$ & $0.6980_{-0.4777}^{+0.5100}$ & -- & $4.163_{-2.509}^{+2.622}$ & $0.04797_{-0.03336}^{+0.03360}$ & $19.17_{-8.35}^{+8.19}$ & $2.092_{-0.801}^{+0.845}$ \\
\hline
\multicolumn{8}{c}{\bf MGWP Signal Injection} \\
\hline
Injected Values & -- & -- & $0.5$ & $3.5$ & $0.03$ & $17$ & $3$ \\
\hline
CE+ET & -- & -- & $0.8632_{-0.2793}^{+0.2658}$ & $3.119_{-0.147}^{+0.144}$ & $0.02807_{-0.00315}^{+0.00324}$ & $19.72_{-1.02}^{+1.04}$ & $2.306_{-0.384}^{+0.397}$ \\
CE & -- & -- & $1.0610_{-0.3129}^{+0.3050}$ & $3.106_{-0.169}^{+0.161}$ & $0.03020_{-0.00429}^{+0.00438}$ & $19.62_{-1.18}^{+1.26}$ & $2.067_{-0.433}^{+0.424}$ \\
ET & -- & -- & $0.2885_{-0.7697}^{+0.7405}$ & $3.749_{-0.894}^{+0.966}$ & $0.05571_{-0.01982}^{+0.01875}$ & $26.87_{-7.25}^{+7.52}$ & $2.404_{-0.864}^{+0.823}$ \\
LIGO & -- & -- & $0.1637_{-0.7897}^{+0.8403}$ & $3.973_{-2.747}^{+2.891}$ & $0.05243_{-0.03129}^{+0.03093}$ & $21.86_{-10.57}^{+11.62}$ & $2.559_{-1.024}^{+1.077}$ \\
\hline
\multicolumn{8}{c}{\bf ppE + MGWP Signal Injection} \\
\hline
Injected Values & $4$ & $0.5$ & $0.5$ & $3.5$ & $0.03$ & $17$ & $3$ \\
\hline
CE+ET & $3.589_{-0.429}^{+0.422}$ & $0.5762_{-0.1707}^{+0.1744}$ & $1.2840_{-0.2830}^{+0.2790}$ & $2.939_{-0.110}^{+0.116}$ & $0.03625_{-0.00685}^{+0.00689}$ & $21.24_{-1.09}^{+1.08}$ & $1.615_{-0.427}^{+0.438}$ \\
CE & $3.714_{-0.511}^{+0.579}$ & $0.6412_{-0.2263}^{+0.2477}$ & $1.2880_{-0.3102}^{+0.2950}$ & $3.006_{-0.135}^{+0.145}$ & $0.03879_{-0.00770}^{+0.00787}$ & $20.71_{-1.39}^{+1.35}$ & $1.656_{-0.450}^{+0.503}$ \\
ET & $4.015_{-0.680}^{+0.694}$ & $0.6206_{-0.3627}^{+0.3601}$ & $0.3728_{-0.7358}^{+0.7432}$ & $3.834_{-0.934}^{+0.937}$ & $0.06260_{-0.01902}^{+0.02020}$ & $27.59_{-7.20}^{+7.50}$ & $2.228_{-0.799}^{+0.784}$ \\
LIGO & $2.885_{-1.563}^{+1.459}$ & $0.6149_{-0.4621}^{+0.5091}$ & $0.2187_{-0.8586}^{+0.8803}$ & $4.126_{-2.614}^{+2.516}$ & $0.05133_{-0.03029}^{+0.03199}$ & $21.29_{-11.12}^{+11.79}$ & $2.441_{-0.961}^{+0.960}$ \\
\hline
\end{tabular}
\end{table*}

\begin{table*}[ht]
\centering
\caption{Relative uncertainty (precision) of the inferred parameters.}
\label{tab:precision2}
\renewcommand{\arraystretch}{1.3}
\begin{tabular}{lccccccc}
\hline
Detector & $a$ & $\alpha_\mathrm{ppE}$ & $\alpha_\mathrm{M0}$ & $\alpha_\mathrm{IMF}$ & $\lambda_\mathrm{peak}$ & $\mathcal{R}$ & $\gamma$ \\
\hline
\multicolumn{8}{c}{\bf GR Signal Injection} \\
\hline
CE+ET & 28.5\% & 81.1\% & -- & 4.4\% & 10.1\% & 6.5\% & 4.3\% \\
CE & 32.2\% & 76.7\% & -- & 6.0\% & 14.0\% & 8.6\% & 5.6\% \\
ET & 49.7\% & 70.1\% & -- & 36.2\% & 31.6\% & 27.7\% & 30.2\% \\
LIGO & 64.0\% & 66.8\% & -- & 65.0\% & 64.4\% & 44.5\% & 40.4\% \\
\hline
\multicolumn{8}{c}{\bf ppE Signal Injection} \\
\hline
CE+ET & 15.8\% & 39.6\% & -- & 4.3\% & 18.3\% & 6.2\% & 4.1\% \\
CE & 18.2\% & 44.7\% & -- & 6.2\% & 20.0\% & 9.0\% & 6.1\% \\
ET & 17.0\% & 57.8\% & -- & 34.9\% & 34.0\% & 27.6\% & 29.3\% \\
LIGO & 48.6\% & 70.8\% & -- & 61.6\% & 69.8\% & 43.1\% & 39.3\% \\
\hline
\multicolumn{8}{c}{\bf MGWP Signal Injection} \\
\hline
CE+ET & -- & -- & 31.6\% & 4.7\% & 11.4\% & 5.2\% & 16.9\% \\
CE & -- & -- & 29.1\% & 5.3\% & 14.4\% & 6.2\% & 20.7\% \\
ET & -- & -- & 261.7\% & 24.8\% & 34.6\% & 27.5\% & 35.1\% \\
LIGO & -- & -- & 497.9\% & 71.0\% & 59.3\% & 50.8\% & 41.1\% \\
\hline
\multicolumn{8}{c}{\bf ppE + MGWP Signal Injection} \\
\hline
CE+ET & 11.9\% & 29.9\% & 21.9\% & 3.8\% & 19.0\% & 5.1\% & 26.8\% \\
CE & 14.7\% & 37.0\% & 23.5\% & 4.7\% & 20.1\% & 6.6\% & 28.8\% \\
ET & 17.1\% & 58.2\% & 198.4\% & 24.4\% & 31.3\% & 26.6\% & 35.5\% \\
LIGO & 52.4\% & 79.0\% & 397.6\% & 62.2\% & 60.7\% & 53.8\% & 39.3\% \\
\hline
\end{tabular}
\end{table*}

\begin{table*}[ht]
\centering
\caption{Relative uncertainties computed with respect to the injected (fiducial) parameter values.}
\label{tab:precision_fiducial}
\renewcommand{\arraystretch}{1.3}
\begin{tabular}{lccccccc}
\hline
Detector & $a$ & $\alpha_\mathrm{ppE}$ & $\alpha_\mathrm{M0}$ & $\alpha_\mathrm{IMF}$ & $\lambda_\mathrm{peak}$ & $\mathcal{R}$ & $\gamma$ \\
\hline
\multicolumn{8}{c}{\bf GR Signal Injection} \\
\hline
CE+ET & 25.4\% & -- & -- & 4.1\% & 11.4\% & 7.4\% & 3.9\% \\
CE & 27.5\% & -- & -- & 5.6\% & 16.3\% & 9.9\% & 5.1\% \\
ET & 35.9\% & -- & -- & 41.8\% & 69.2\% & 42.9\% & 20.3\% \\
LIGO & 40.8\% & -- & -- & 73.3\% & 104.9\% & 52.0\% & 28.0\% \\
\hline
\multicolumn{8}{c}{\bf ppE Signal Injection} \\
\hline
CE+ET & 15.9\% & 42.2\% & -- & 4.1\% & 19.5\% & 7.0\% & 3.8\% \\
CE & 18.0\% & 46.1\% & -- & 5.9\% & 22.2\% & 10.1\% & 5.7\% \\
ET & 17.3\% & 79.2\% & -- & 41.5\% & 74.7\% & 42.5\% & 19.9\% \\
LIGO & 35.9\% & 98.8\% & -- & 73.3\% & 111.6\% & 48.6\% & 27.4\% \\
\hline
\multicolumn{8}{c}{\bf MGWP Signal Injection} \\
\hline
CE+ET & -- & -- & 54.5\% & 4.2\% & 10.6\% & 6.1\% & 13.0\% \\
CE & -- & -- & 61.8\% & 4.7\% & 14.5\% & 7.2\% & 14.3\% \\
ET & -- & -- & 151.0\% & 26.6\% & 64.3\% & 43.4\% & 28.1\% \\
LIGO & -- & -- & 163.0\% & 80.5\% & 103.7\% & 65.3\% & 35.0\% \\
\hline
\multicolumn{8}{c}{\bf ppE + MGWP Signal Injection} \\
\hline
CE+ET & 10.6\% & 34.5\% & 56.2\% & 3.2\% & 22.9\% & 6.4\% & 14.4\% \\
CE & 13.6\% & 47.4\% & 60.5\% & 4.0\% & 25.9\% & 8.1\% & 15.9\% \\
ET & 17.2\% & 72.3\% & 147.9\% & 26.7\% & 65.4\% & 43.2\% & 26.4\% \\
LIGO & 37.8\% & 97.1\% & 173.9\% & 73.3\% & 103.8\% & 67.4\% & 32.0\% \\
\hline
\end{tabular}
\end{table*}

{The forecast constraints on the modified-gravity parameters and the main population hyperparameters are collected in Tables~\ref{tab:inf_geral}, \ref{tab:precision2} and~\ref{tab:precision_fiducial}, which present three complementary views of the same injection--recovery analyses. We consider four injections: a pure GR signal; a ppE waveform-generation signal~; a modified GW propagation (MGWP) signal parametrized by; and a joint ppE+MGWP signal in which both classes of deviation are present simultaneously. Each injection is analyzed with four detector configurations: the combined CE+ET network, CE, ET, and Advanced LIGO.}

Table~\ref{tab:inf_geral} reports the injected (fiducial) values together with the recovered posteriors, quoted as median values with $1\sigma$ credible intervals for each configuration. The remaining tables summarize the associated relative uncertainties under two different normalizations: Table~\ref{tab:precision2} divides the averaged error bar by the recovered central value, $100\times[(\sigma_+ + \sigma_-)/2]/\theta_{\rm rec}$, and thus measures the intrinsic precision with which each parameter is determined, whereas Table~\ref{tab:precision_fiducial} divides it by the injected fiducial value, $100\times[(\sigma_+ + \sigma_-)/2]/\theta_{\rm inj}$, quantifying the accuracy with which the true input is recovered. In the GR injection the latter normalization is undefined for $\alpha_{\rm ppE}$, whose fiducial value vanishes, and the corresponding entries are left blank.

Overall, the results show a clear hierarchy in constraining power among the individual detector configurations. Cosmic Explorer (CE) provides the tightest constraints in nearly all cases, followed by the Einstein Telescope (ET), while Advanced LIGO exhibits substantially weaker performance due to its reduced sensitivity to the SGWB signal. This trend holds for the astrophysical population quantities in all four injections, and for the modified-gravity parameters whenever a non-zero deviation from GR is injected. The one exception among the latter is the PN exponent $a$ in the ppE injection, for which the two third-generation detectors perform comparably, ET reaching $17.0\%$ against $18.2\%$ for CE in Table~\ref{tab:precision2}. The hierarchy also does not apply to $\alpha_{\rm ppE}$ in the GR injection: since its fiducial value vanishes, the posterior is prior-dominated and the error bar is nearly identical for all configurations, so the normalization of Table~\ref{tab:precision2} no longer provides a meaningful ranking.

{The combined CE+ET configuration provides the tightest constraints of all the setups considered here, and the pattern of improvement is itself informative. For the parameters constrained through the spectral shape of $\Omega_{\rm GW}(f)$ -- in particular the PN exponent $a$ -- the gain over CE alone is systematic but moderate, the relative uncertainty decreasing from $18.2\%$ to $15.8\%$ in the ppE injection and from $14.7\%$ to $11.9\%$ in the joint one. The same holds for the ppE amplitude, which improves from $44.7\%$ to $39.6\%$. For the amplitude-like propagation parameter $\alpha_{M0}$ the picture depends on the normalization adopted: in the MGWP injection the combination appears marginally less precise than CE ($31.6\%$ against $29.1\%$ in Table~\ref{tab:precision2}) while being more accurate ($54.5\%$ against $61.8\%$ in Table~\ref{tab:precision_fiducial}). The two statements are not in conflict: the absolute credible interval is in fact narrower for CE+ET, and the apparent gain in precision of CE alone follows from its recovered median being displaced further above the injected value, which inflates the denominator of Table~\ref{tab:precision2}. That the improvement is moderate rather than dramatic follows from the frequency dependence of the combined network response: the CE+ET baseline retains most of its coherence at the lowest frequencies~\cite{Allen:1997ad}, where modified propagation acts as a nearly scale-independent amplitude rescaling, while the corresponding overlap reduction function decorrelates in the $\sim 10^2$~Hz region that carries much of the spectral-shape information relevant for the ppE sector.}

\subsection{Constraints on ppE waveform modifications}

For the GR signal injection, corresponding to $\alpha_{\rm ppE}=0$, the recovered posterior distributions remain statistically consistent with the absence of modified-gravity effects. In particular, the inferred values of $\alpha_{\rm ppE}$ are centered close to zero for all detector configurations, indicating that the inference pipeline does not introduce significant biases when recovering a pure GR signal. {In this regime, the PN exponent $a$ remains only weakly constrained, with relative uncertainties of $28.5\%$ for CE+ET and ranging from approximately $30\%$--$50\%$ for CE and ET, reaching $64\%$ for LIGO.} Its recovered medians lie systematically below the injected value, by an amount that decreases as the sensitivity improves and that remains within the $1\sigma$ credible interval in every configuration. This behavior is expected, since for $\alpha_{\rm ppE}\simeq0$ the SGWB becomes only weakly sensitive to the frequency-dependent ppE correction.

For the non-GR ppE injection, with fiducial values $(a,\alpha_{\rm ppE})=(4,0.5)$, both CE and ET successfully recover the injected parameters within the inferred credible intervals. Among the individual configurations, CE provides the most accurate reconstruction, yielding
\begin{equation}
a = 3.955_{-0.699}^{+0.742},
\qquad
\alpha_{\rm ppE}=0.5149_{-0.2309}^{+0.2296},
\end{equation}
demonstrating that third-generation detectors are capable of probing frequency-dependent beyond-GR effects in the SGWB.

{The combined CE+ET configuration recovers the injected parameters within $1\sigma$ and with the smallest uncertainties among all the setups considered, $a = 4.005^{+0.647}_{-0.621}$ and $\alpha_{\rm ppE} = 0.5335^{+0.2172}_{-0.2052}$.}

\begin{figure*}[t]
\centering
\includegraphics[width=0.45\textwidth]{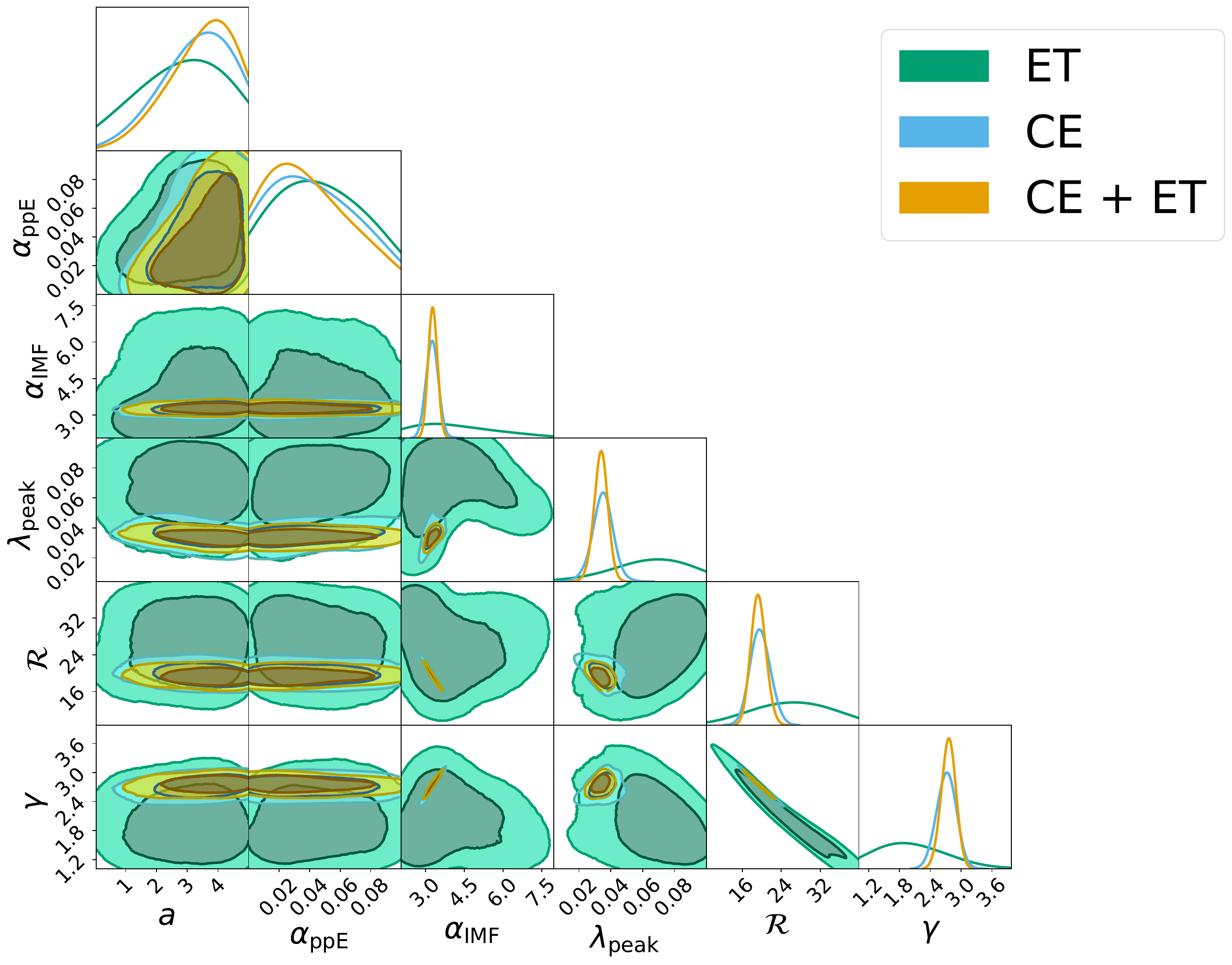}
\hfill
\includegraphics[width=0.45\textwidth]{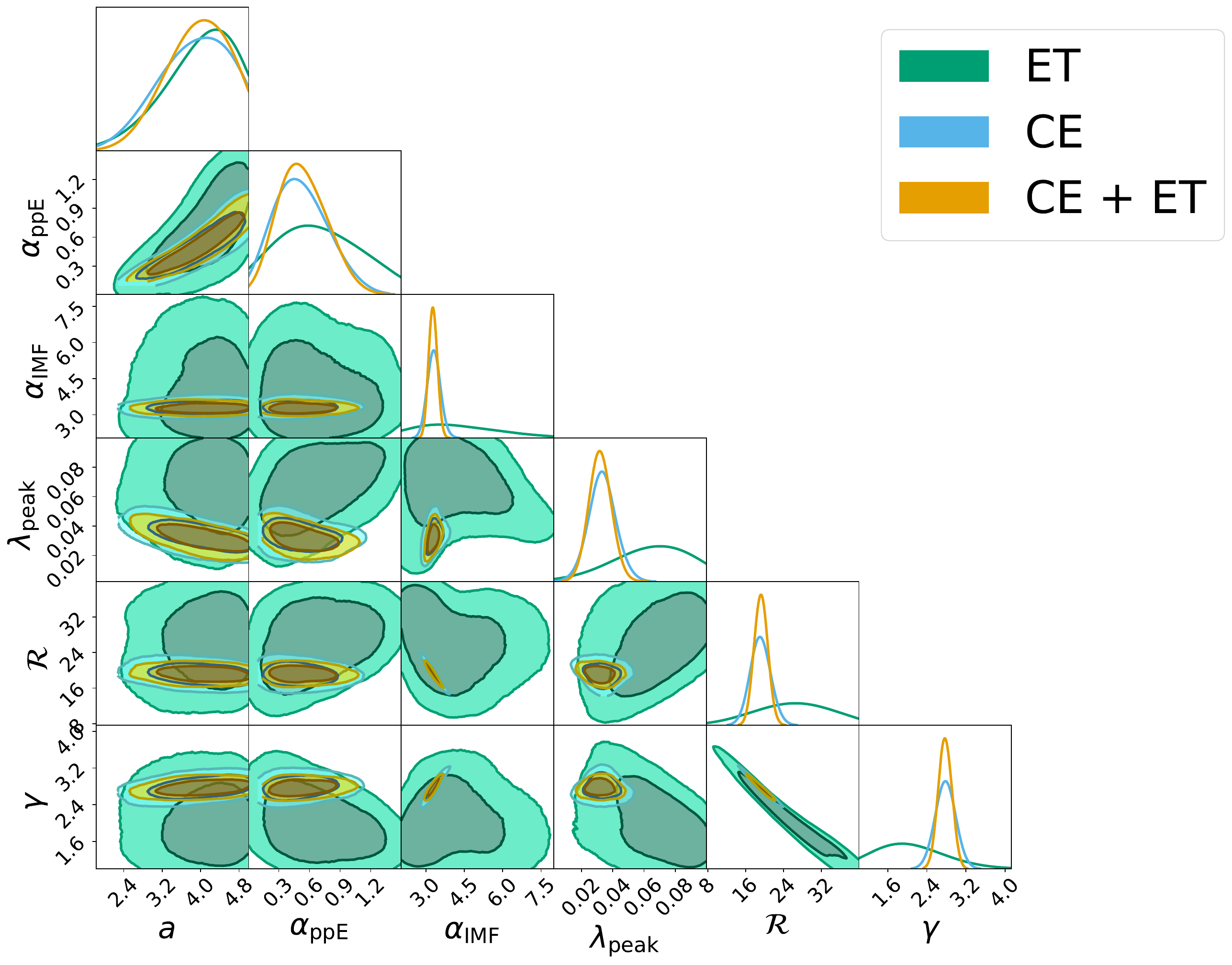}
\caption{
Posterior distributions for the ppE amplitude parameter $\alpha_{\rm ppE}$, the mass-function slope $\alpha_{\rm IMF}$, the Gaussian peak fraction $\lambda_{\rm peak}$, the local merger rate $\mathcal{R}$, and the ppE PN exponent $a$, obtained for ET, CE and the combined CE+ET network.
\textit{Left panel:} GR injection with $\alpha_{\rm ppE}=0$.
\textit{Right panel:} ppE injection with $\alpha_{\rm ppE}=0.5$ and $a=4$.
The contours correspond to the $68\%$ and $95\%$ credible regions.
}
\label{fig:PS_ppE_case}
\end{figure*}

Figure~\ref{fig:PS_ppE_case} shows the posterior distributions for the ppE parameters and BBH population hyperparameters obtained with ET, CE {and CE+ET} for both the GR and ppE injections. In the GR injection case ($\alpha_{\rm ppE}=0$; left panel), the posterior distribution of $\alpha_{\rm ppE}$ remains consistent with zero for all configurations, indicating no evidence for deviations from GR. In contrast, for the ppE injection
($\alpha_{\rm ppE}=0.5$, $a=4$; right panel), the posterior shifts away from zero, demonstrating that the three configurations are able to recover the injected modified-gravity signal.

The figure also reveals correlations and parameter degeneracies. The strongest degeneracy appears between $\alpha_{\rm ppE}$ and the BBH population parameters, particularly $\alpha_{\rm IMF}$ and $\lambda_{\rm peak}$. This behavior arises because changes in the ppE amplitude can partially mimic variations in the intrinsic BBH mass distribution, modifying both the overall SGWB amplitude and its spectral shape. A noticeable correlation is also observed between $\alpha_{\rm ppE}$ and the PN exponent $a$, especially in the modified-gravity injection case, reflecting the fact that both parameters jointly determine the frequency dependence of the ppE correction.

Additionally, the local merger rate $\mathcal{R}$ exhibits partial degeneracy with $\alpha_{\rm ppE}$, since both parameters affect the normalization of the SGWB spectrum. The parameter $a$ is significantly better constrained in the non-GR injection compared to the GR case. This improvement arises because the injected ppE correction introduces a measurable frequency-dependent distortion in the SGWB spectrum, allowing the shape information of $\Omega_{\rm GW}(f)$ to break degeneracies between amplitude and spectral parameters. The relative uncertainty on $a$ decreases to $\sim18\%$ for CE and $\sim17\%$ for ET, while LIGO remains unable to accurately reconstruct the injected signal due to its substantially larger uncertainties.

By contrast, the amplitude parameter $\alpha_{\rm ppE}$ remains comparatively weakly constrained even for third-generation detectors, with relative uncertainties at the level of $\sim40\%$--$60\%$. This reflects residual degeneracies between the overall SGWB normalization and astrophysical population parameters, particularly the merger-rate amplitude $\mathcal{R}$ and the BBH mass distribution.

\begin{figure}[t]
\centering
\includegraphics[width=0.45\textwidth]{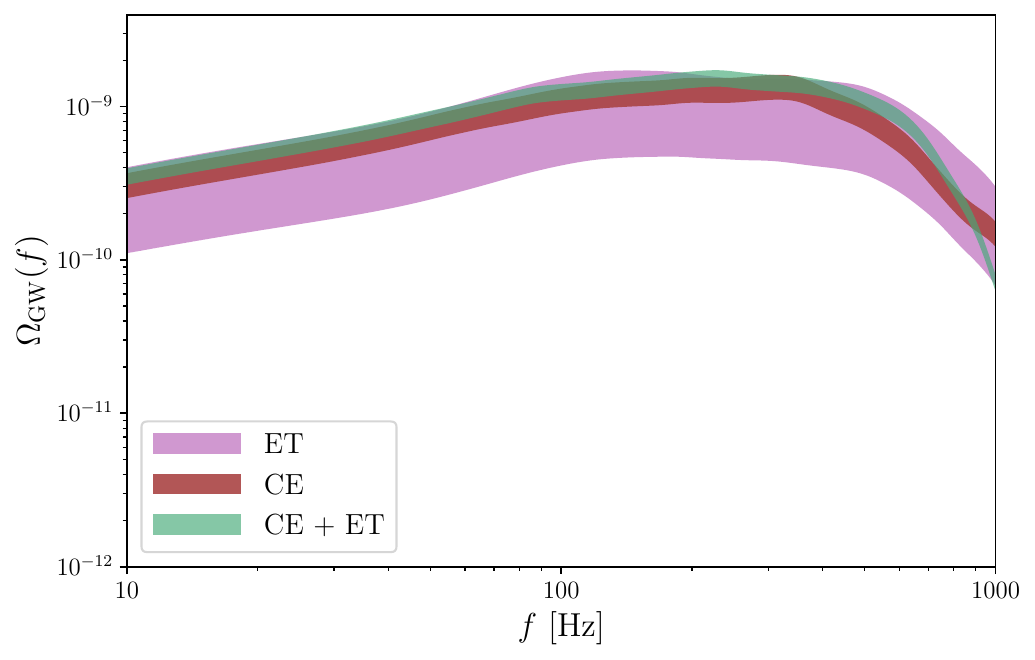}
\caption{
Reconstructed SGWB energy-density spectrum $\Omega_{\rm GW}(f)$ at 2$\sigma$ credible
interval for the ppE injection--recovery analysis performed with ET (pink), CE (dark red)
and the combined CE+ET network (green). The figure shows the non-GR injection case with
$\alpha_{\rm ppE}=0.5$ and $a=4$.
}
\label{fig:GWs_reconstrution}
\end{figure}

Figure~\ref{fig:GWs_reconstrution} shows the reconstructed SGWB energy-density spectrum $\Omega_{\rm GW}(f)$ obtained from the ppE injection--recovery analysis for ET, CE {and CE+ET} in the non-GR case with $\alpha_{\rm ppE}=0.5$ and $a=4$. All three configurations successfully recover the injected signal within the $2\sigma$ credible region over the full frequency range. The tightest constraints are achieved around $\sim100$--$300\, {\rm Hz}$, where the detector sensitivity is maximal, while broader posteriors at low and high frequencies reflect reduced sensitivity and residual parameter degeneracies. CE provides systematically tighter constraints than ET, indicating a stronger capability to reconstruct the amplitude and spectral shape of modified-gravity SGWB signals.

It is important to emphasize that the dominant uncertainties in the non-GR parameter reconstruction originate from the BBH population hyperparameters. Strong degeneracies exist between the modified-gravity parameters and the astrophysical population model, particularly because both can affect the overall amplitude and spectral shape of the SGWB. Consequently, improved observational constraints or stronger theoretical priors on the BBH population could significantly enhance the accuracy of beyond-GR measurements by partially breaking these correlations and reducing the allowed parameter volume.

In the present work, however, we intentionally adopt a conservative and fully general forecasting framework in which all relevant population hyperparameters are simultaneously varied. This approach avoids artificially optimistic constraints and provides a more realistic assessment of the capability of future detectors to probe modified gravity through the SGWB. Exploring scenarios with partially fixed or externally constrained population parameters may further improve the sensitivity to non-GR effects and will be investigated in future work.

\subsection{Constraints on modified GW propagation}
\label{constrain_mgwp}

The cosmological analysis exhibits a qualitatively different behavior compared to the ppE scenario. Since modified GW propagation primarily acts as a smooth redshift-dependent rescaling of the SGWB amplitude, rather than introducing a strong frequency-dependent distortion, the parameter $\alpha_{M0}$ is substantially more degenerate with astrophysical uncertainties.

For the non-GR cosmological injection, corresponding to $\alpha_{M0}=0.5$, the two third-generation detectors behave rather differently. ET yields
\begin{equation}
\alpha_{M0}=0.2885_{-0.7697}^{+0.7405},
\end{equation}
consistent with the injected value but with an uncertainty large enough for the posterior to be essentially uninformative, while CE obtains a much narrower constraint,
\begin{equation}
\alpha_{M0}=1.0610_{-0.3129}^{+0.3050},
\end{equation}
whose median is displaced upward with respect to the injected value.

{The combined CE+ET network provides the tightest bound on the propagation parameter,
\begin{equation}
\alpha_{M0} = 0.8632^{+0.2658}_{-0.2793},
\end{equation}
corresponding to a relative uncertainty of $\sim 31\%$, roughly a factor of two better than
ET and comparable to CE. The recovered median is displaced upward by about $1\sigma$ with
respect to the injected value, and the improvement is accompanied by a poorly recovered
mass-function slope $\alpha_{\rm IMF}$, indicating that part of the gain in $\alpha_{M0}$ is
obtained at the expense of a residual degeneracy transferred to the mass-function sector.}

However, for the individual detector configurations the relative uncertainty on $\alpha_{M0}$ remains large, reaching $\sim 60\%$ for CE and exceeding $150\%$ for ET, even for third-generation observatories. This behavior reflects the intrinsic nature of modified GW propagation effects, which primarily alter the overall SGWB amplitude through a cumulative cosmological reweighting of distant sources. As a consequence, the signal becomes strongly degenerate with the merger-rate evolution and the normalization of the BBH population. This interpretation is further supported by the behavior of the astrophysical hyperparameters. In the cosmological analysis, the low-redshift evolution index $\gamma$ exhibits noticeably larger uncertainties than in the ppE case, highlighting the strong correlation between modified GW propagation effects and the redshift evolution of the BBH population.

Therefore, improving the constraints on $\alpha_{M0}$ would require the same strategy discussed previously for the ppE analysis. In particular, tighter observational measurements or stronger theoretical priors on the BBH population and merger-rate evolution could substantially reduce these degeneracies, leading to significantly improved bounds on modified GW propagation scenarios.

\begin{figure*}[t]
\centering
\includegraphics[width=0.45\textwidth]{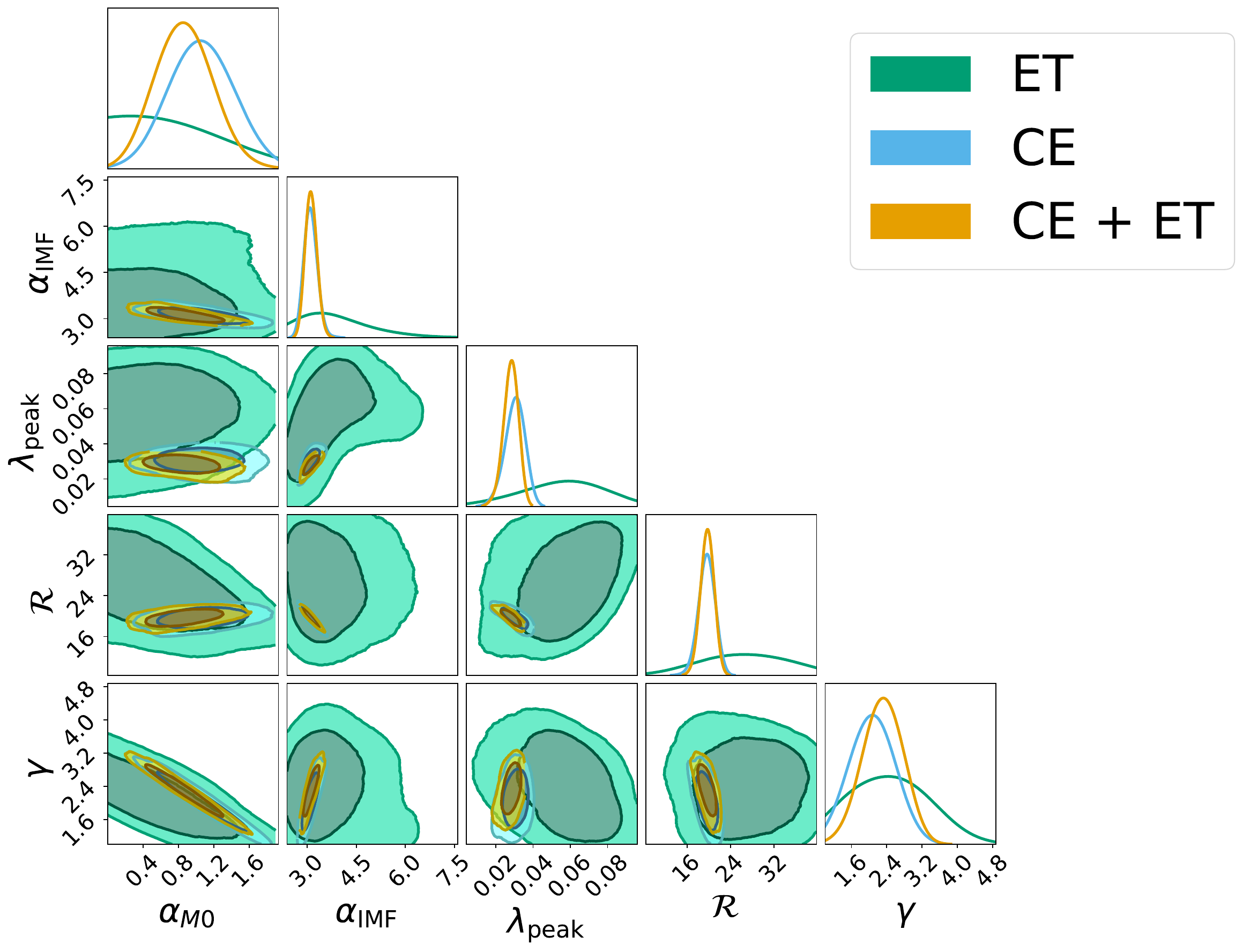} \,\,\,
\includegraphics[width=0.45\textwidth]{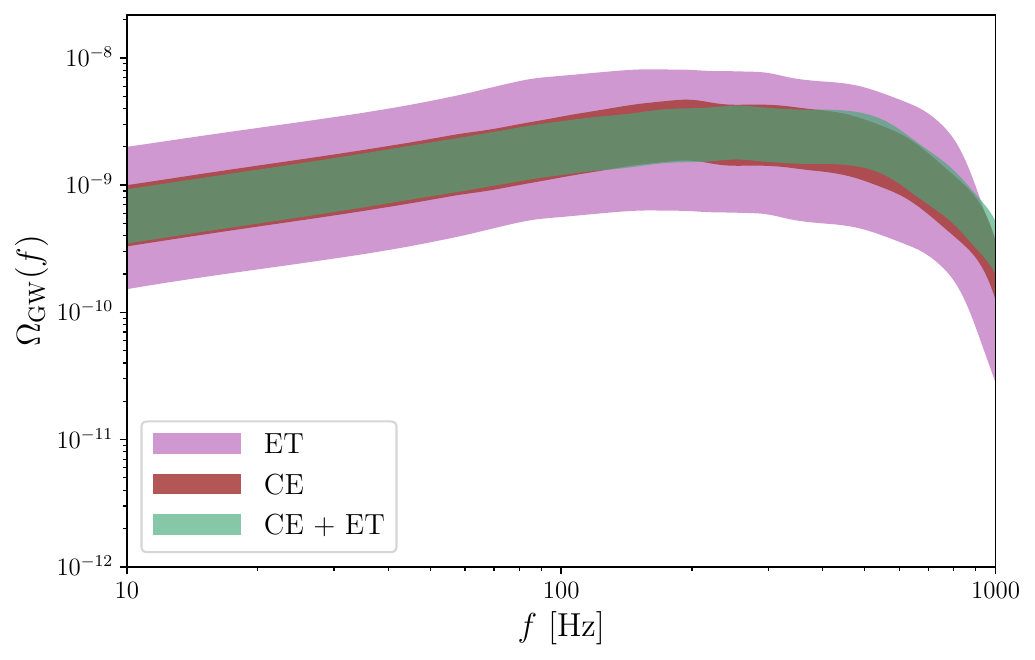} 
\caption{
Left panel: Posterior distributions for the running Planck-mass amplitude $\alpha_{M0}$, the
mass-function slope $\alpha_{\rm IMF}$, the Gaussian peak fraction $\lambda_{\rm peak}$, the
local merger rate $\mathcal{R}$, and the low-redshift evolution index $\gamma$, obtained
with ET, CE and the combined CE+ET network for the non-GR injection with $\alpha_{M0}=0.5$.
The contours correspond to the $68\%$ and $95\%$ credible regions.
Right panel: Reconstructed SGWB energy-density spectrum $\Omega_{\rm GW}(f)$ for the
cosmological injection--recovery analysis. The reconstructed spectra are consistent with the
injected signal within the inferred credible intervals.
}
\label{fig:cosmology_case}
\end{figure*}

\begin{figure*}[t]
\centering
\includegraphics[width=0.50\textwidth]{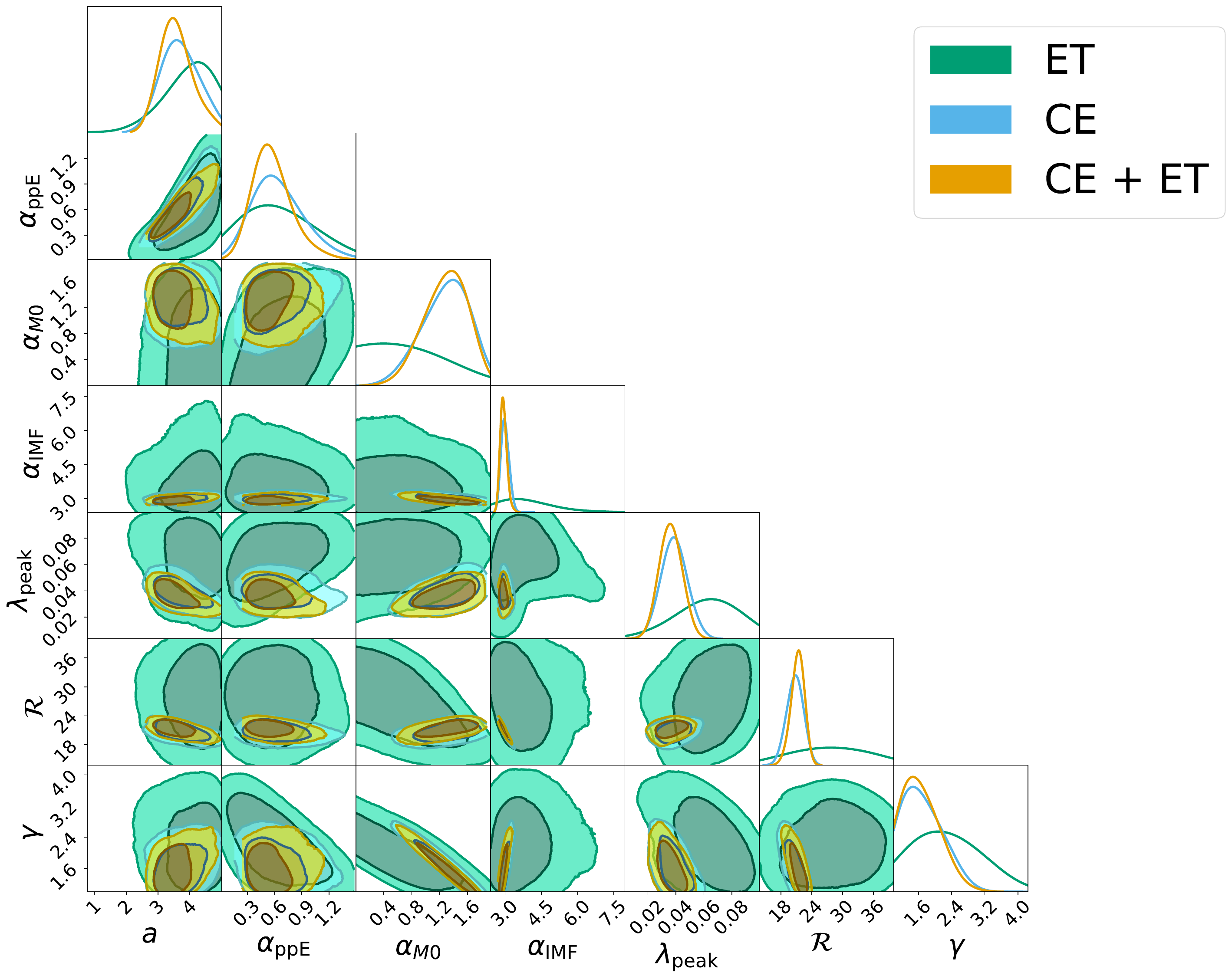} \,\,\,
\includegraphics[width=0.40\textwidth]{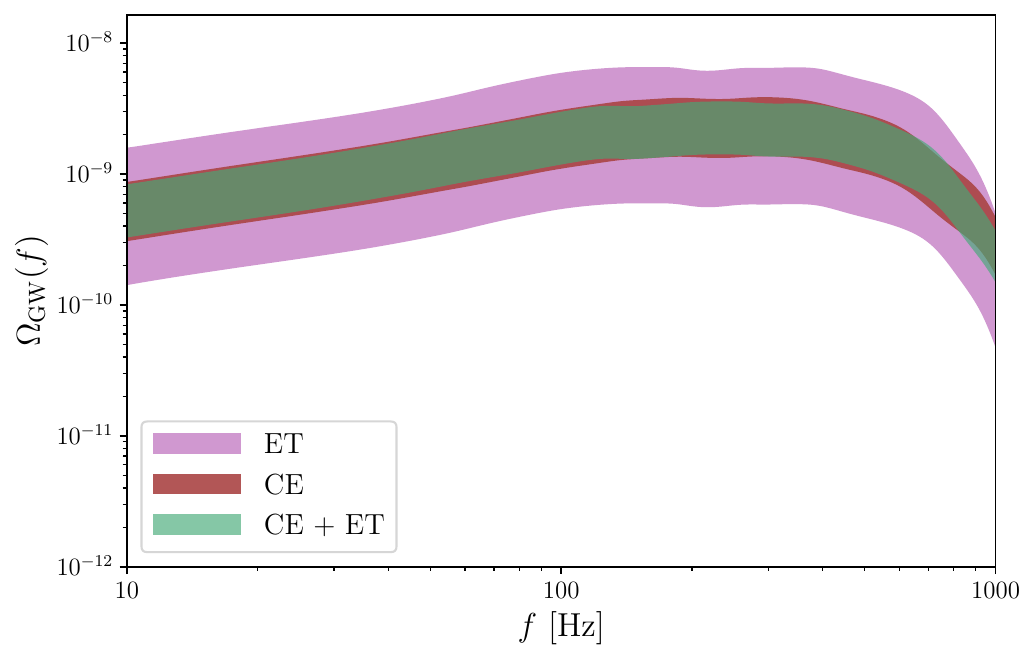} 
\caption{Left panel: Posterior distributions for the ppE parameters ($a$ and
$\alpha_{\text{ppE}}$), the running Planck-mass parameter
$\alpha_{M_0}$, the mass-function slope
$\alpha_{\text{IMF}}$, the Gaussian peak fraction $\lambda_{\text{peak}}$, the local merger
rate $\mathcal{R}$, and the low-redshift evolution index $\gamma$, obtained with ET, CE and
the combined CE+ET network for the combined ppE + MGWP injection with
$\alpha_{\text{ppE}} = 0.5$, $a = 4$, and $\alpha_{M_0} = 0.5$. The contours correspond to
the 68\% and 95\% credible regions. Right panel: Reconstructed SGWB energy-density spectrum
$\Omega_{\text{GW}}(f)$ for the joint ppE + MGWP injection--recovery analysis. The
reconstructed spectra are consistent with the injected signal within the inferred credible
intervals, demonstrating the capability of the three configurations to simultaneously
constrain beyond-GR effects in both waveform-generation and propagation sectors.}
\label{fig:cosmologyppE_case}
\end{figure*}

Figure~\ref{fig:cosmology_case} shows the parameter-estimation results for the modified GW propagation scenario with $\alpha_{M0}=0.5$. The left panel presents the posterior distributions for $\alpha_{M0}$ and the BBH population hyperparameters obtained with ET, CE {and CE+ET}. The injected value lies within the credible regions of all three configurations, although significant degeneracies are observed with the merger-rate parameters $\mathcal{R}$ and $\gamma$. The right panel shows the reconstructed SGWB energy-density spectrum $\Omega_{\rm GW}(f)$, demonstrating that the injected signal is recovered within the inferred credible intervals. CE provides substantially tighter constraints than ET over most of the observable frequency range.

\subsection{Joint ppE + MGWP analysis}
\label{sec:joint}

{The two classes of beyond-GR effects considered above are not mutually exclusive: a given modified-gravity theory may alter both the waveform generated during the inspiral and the propagation of the signal over cosmological distances~\cite{Nishizawa:2017nef,Tahura:2018zuq}. To assess whether the SGWB can disentangle the two sectors, we perform an additional injection in which both deviations are present simultaneously, with fiducial values $(a, \alpha_{\rm ppE}, \alpha_{M0}) = (4,\,0.5,\,0.5)$, recovering the modified-gravity  parameters jointly with the population hyperparameters.}

{Figure~\ref{fig:cosmologyppE_case} shows the results of this analysis. The left panel displays the posterior distributions obtained with ET, CE and CE+ET, while the right panel shows the corresponding reconstructed SGWB energy-density spectrum. The three configurations recover the injected signal within the inferred credible intervals, demonstrating that the two beyond-GR sectors can be constrained simultaneously rather than one at a time. For CE we
obtain}
\begin{align}
a &= 3.714^{+0.579}_{-0.511}, \nonumber\\
\alpha_{\rm ppE} &= 0.6412^{+0.2477}_{-0.2263}, \nonumber\\
\alpha_{M0} &= 1.2880^{+0.2950}_{-0.3102},
\label{eq:joint_CE}
\end{align}
{in agreement with the injected values in the waveform-generation sector.}

{The comparison with the individual analyses is instructive. The spectral-shape parameter $a$ is essentially unaffected by opening the propagation sector: its relative uncertainty remains at $\sim 17\%$ for ET and even improves from $18.2\%$ to $14.7\%$ for CE, as reported in Table~\ref{tab:precision2}. This robustness is a direct consequence of the fact that $\alpha_{M0}$ enters $\Omega_{\rm GW}(f)$ as a nearly scale-independent rescaling, Eq.~\eqref{eq:OmegaGW_effective_theta}, and therefore cannot mimic the frequency-dependent
distortion controlled by $a$, Eq.~\eqref{eq:OmegaGW_ppE_final}. The amplitude-like parameters, in contrast, pay a price for being varied together: although their credible intervals do not widen, their recovered medians move further away from the injected values, $\alpha_{\rm ppE}$ shifting from $0.5149$ to $0.6412$ and $\alpha_{M0}$ from $1.0610$ to $1.2880$ for CE, since both parameters can shift the overall normalization of the spectrum and the degeneracy between them is only partially broken by the different frequency behavior of the two corrections.}

{Advanced LIGO is unable to constrain the propagation sector in this configuration: the recovered posterior, $\alpha_{M0} = 0.2187^{+0.8803}_{-0.8586}$, is essentially prior-dominated. This is consistent with the picture emerging from the individual analyses, in which second-generation sensitivity is insufficient to resolve amplitude-level deviations from GR once the astrophysical population is marginalized over.}

{It is worth stressing that the same two sectors are probed in a qualitatively different way by resolved events and by the background. Parametrized waveform tests and modified-propagation analyses of the GWTC catalogs currently provide the leading bounds on these deviations~\cite{LIGOScientific:2026fcf,LIGOScientific:2025jau,LIGOScientific:2026uyd}, but they are dominated by the loudest, low-redshift events. The SGWB, being an incoherent superposition of the full unresolved population, probes the same parameters integrated over cosmic history~\cite{Nunes:2020rmr,Ezquiaga:2021ayr}, and the joint analysis presented here shows that this integrated information is sufficient to constrain both sectors at once with third-generation detectors.}

\subsection{Astrophysical population constraints}

The astrophysical hyperparameters exhibit different levels of recoverability depending on how strongly they affect the SGWB amplitude and spectral shape. In general, the local merger rate $\mathcal{R}$ and the low-redshift evolution parameter $\gamma$ are among the best constrained astrophysical quantities, particularly for third-generation detectors. In the ppE analysis, CE achieves uncertainties of approximately $9\%$ for $\mathcal{R}$ and $6\%$ for $\gamma$, demonstrating that SGWB observations contain significant information about the global BBH merger history.

The IMF slope $\alpha_{\rm IMF}$ is well constrained by CE, with uncertainties of a few per cent, degrading to $\sim25\%$--$35\%$ for ET. Since this parameter controls the abundance of massive BBHs, it directly impacts both the normalization of the SGWB spectrum and the position of the high-frequency turnover. {By contrast, the Gaussian-peak fraction $\lambda_{\rm peak}$ is systematically less well determined, with uncertainties of $\sim10\%$--$20\%$ for CE and CE+ET, $\sim30\%$ for ET, and approaching or exceeding $100\%$ for LIGO.} This indicates that the SGWB is comparatively insensitive to subdominant features of the BBH mass distribution once marginalized over merger-rate and modified-gravity uncertainties.

The results obtained in the modified-gravity analyses further highlight the strong interplay between astrophysical and beyond-GR parameters. In the ppE framework, the frequency-dependent correction introduces characteristic distortions in the SGWB spectral shape, allowing the parameter $a$ to be relatively well constrained once a nonzero signal is injected. Nevertheless, significant degeneracies remain between $\alpha_{\rm ppE}$ and astrophysical quantities such as $\mathcal{R}$, $\alpha_{\rm IMF}$, and $\lambda_{\rm peak}$, since all of them contribute to the overall SGWB amplitude and spectral evolution.

A similar but even stronger effect is observed in the modified GW propagation scenario. In this case, the parameter $\alpha_{M0}$ primarily changes the cumulative cosmological amplitude of the SGWB, making it highly degenerate with the merger-rate evolution parameters $\mathcal{R}$ and $\gamma$. Consequently, the uncertainty on the low-redshift evolution index $\gamma$ in the cosmological analysis becomes noticeably larger than in the ppE case, reflecting the difficulty of separating propagation effects from the intrinsic redshift evolution of the BBH population.

\subsection{Pipeline validation}
\label{sec:validation}

Overall, the results demonstrate that the inference pipeline is internally consistent and capable of robustly recovering the injected SGWB signals within the expected statistical uncertainties. GR injections in both the ppE and modified GW propagation analyses are recovered with posterior distributions statistically compatible with the null modified-gravity hypothesis, indicating that the framework does not introduce artificial deviations from GR.

The detector hierarchy observed throughout the analysis, {CE+ET $\gtrsim$ CE $>$ ET $\gg$ LIGO, is also physically consistent with the expected instrumental sensitivities, providing an additional validation of the methodology.} Third-generation detectors improve the reconstruction of both astrophysical and beyond-GR parameters, particularly for frequency-dependent ppE corrections.

The inferred parameter degeneracies further support the physical consistency of the analysis. In the ppE scenario, the SGWB acquires characteristic frequency-dependent distortions, allowing the PN exponent $a$ to be constrained through spectral-shape information. By contrast, modified GW propagation mainly produces an overall redshift-dependent amplitude rescaling, leading to stronger degeneracies with astrophysical population parameters such as $\mathcal{R}$ and $\gamma$. These qualitatively distinct behaviors naturally emerge from the inference and are consistent with theoretical expectations.

At the astrophysical level, parameters controlling the SGWB normalization and merger-rate evolution are systematically better constrained than the subdominant Gaussian-peak fraction of the BBH mass distribution. However, the analysis also reveals important limitations. In particular, $\alpha_{M0}$ remains only weakly constrained by ET due to strong correlations with astrophysical evolution parameters, while $\alpha_{\rm ppE}$ still exhibits sizable uncertainties associated with residual normalization degeneracies.{The joint ppE + MGWP injection discussed in Sec.~\ref{sec:joint} provides a further consistency check: the modified-gravity parameters are recovered simultaneously, with the residual degeneracy manifesting itself as a displacement of the amplitude-like parameters and of the merger-rate evolution index rather than as a loss of precision, as expected from their common effect on the SGWB normalization.}

Finally, the forecasts presented here should be interpreted as optimistic but physically motivated forecasts for future SGWB observations. Nevertheless, the successful recovery of injected signals and the expected structure of the inferred correlations provide evidence that the pipeline is numerically stable and capable of capturing the leading SGWB signatures associated with both modified gravity and BBH population effects.

The pipeline developed in this work will be made fully publicly available at \texttt{https://github.com/fsrodrigofraga-hash/SGAGE} upon publication of this work.

\section{Final Remarks}
\label{final}

In this work, we investigated the impact of modified gravity effects on the stochastic gravitational-wave background generated by unresolved binary black hole mergers. We considered two complementary beyond-GR scenarios: frequency-dependent waveform modifications within the ppE framework and modified GW propagation effects parameterized through a running Planck-mass model.
Our results show that frequency-dependent ppE corrections generate the clearest SGWB signatures, allowing third-generation detectors to recover the injected signals with meaningful accuracy, particularly through spectral-shape information. In contrast, modified GW propagation effects mainly rescale the SGWB amplitude, leading to stronger degeneracies with the merger-rate evolution and BBH population normalization. {We have further shown, through a joint injection in which both effects are present simultaneously (Sec.~\ref{sec:joint}), that the two sectors can be constrained together: the frequency-dependent ppE distortion remains well measured even when the propagation amplitude is left free, whereas $\alpha_{\rm ppE}$ and $\alpha_{M0}$ retain a residual degeneracy, inherited from their common effect on the overall SGWB normalization, that displaces their recovered medians.}

{We find that the combined CE+ET network provides the strongest overall constraints, closely followed by CE and then ET, while Advanced LIGO remains significantly less sensitive in the SGWB sector.} The inferred detector hierarchy, parameter correlations, and signal reconstructions are physically consistent with theoretical expectations, validating the robustness of the inference pipeline. The gain of the combined network over CE alone is most apparent in the ppE sector, where the spectral-shape information is decisive, whereas for the propagation parameter the two configurations perform comparably. The analysis also demonstrates that the dominant uncertainties in the beyond-GR sector originate from astrophysical population parameters. Consequently, future improvements in BBH population modeling and external astrophysical constraints will be essential for breaking degeneracies and improving the sensitivity to modified gravity. Another promising direction is to investigate how the combination of SGWB observations spanning different frequency bands and
detectors (e.g., LISA, NANOGrav, Einstein Telescope, Cosmic Explorer, and LIGO) can further tighten constraints on the parameters governing modified gravity models. These developments will be presented as a forthcoming publication. Although the present analysis adopts simplifying assumptions, including non-spinning BBHs and idealized detector noise, the results highlight the strong potential of third-generation SGWB observations as a complementary probe of gravitational physics on cosmological scales. Combining SGWB measurements with resolved compact-binary catalogs, standard sirens, and external cosmological datasets will likely play a central role in future tests of General Relativity.

\section*{Acknowledgements}
\noindent {We thank the referee for their careful reading of the manuscript and for their constructive comments and suggestions, which have helped us improve its clarity and presentation.} R.C.N. acknowledges financial support from the Conselho Nacional de Desenvolvimento Científico e Tecnológico (CNPq, National Council for Scientific and Technological Development) through Grant No. 304306/2022-3, and partial financial support from the Fundação de Amparo à Pesquisa do Estado do Rio Grande do Sul (FAPERGS, Research Support Foundation of the State of Rio Grande do Sul) through Grant No. 23/2551-0000848-3 and Grant No. 25/2551-0002612-1.

\section*{Data Availability}
The main pipeline developed in this work will be made publicly available at
\texttt{https://github.com/fsrodrigofraga-hash/SGAGE} upon publication of
this article. Any additional datasets and derived data products generated
or analyzed during this study will be made available by the corresponding
author upon reasonable request.

\makeatletter
\def\bibsection{\section*{References}}
\makeatother

\bibliography{main}

@article{Yunes:2009ke,
    author = "Yunes, Nicolas and Pretorius, Frans",
    title = "{Fundamental Theoretical Bias in Gravitational Wave Astrophysics and the Parameterized Post-Einsteinian Framework}",
    eprint = "0909.3328",
    archivePrefix = "arXiv",
    primaryClass = "gr-qc",
    doi = "10.1103/PhysRevD.80.122003",
    journal = "Phys. Rev. D",
    volume = "80",
    pages = "122003",
    year = "2009"
}

@article{Cornish:2011ys,
    author = "Cornish, Neil and Sampson, Laura and Yunes, Nicolas and Pretorius, Frans",
    title = "{Gravitational Wave Tests of General Relativity with the Parameterized Post-Einsteinian Framework}",
    eprint = "1105.2088",
    archivePrefix = "arXiv",
    primaryClass = "gr-qc",
    doi = "10.1103/PhysRevD.84.062003",
    journal = "Phys. Rev. D",
    volume = "84",
    pages = "062003",
    year = "2011"
}

@article{Tahura:2018zuq,
    author = "Tahura, Sharaban and Yagi, Kent",
    title = "{Parameterized Post-Einsteinian Gravitational Waveforms in Various Modified Theories of Gravity}",
    eprint = "1809.00259",
    archivePrefix = "arXiv",
    primaryClass = "gr-qc",
    doi = "10.1103/PhysRevD.98.084042",
    journal = "Phys. Rev. D",
    volume = "98",
    number = "8",
    pages = "084042",
    year = "2018",
    note = "[Erratum: Phys.Rev.D 101, 109902 (2020)]"
}

@article{Nishizawa:2017nef,
    author = "Nishizawa, Atsushi",
    title = "{Generalized framework for testing gravity with gravitational-wave propagation. I. Formulation}",
    eprint = "1710.04825",
    archivePrefix = "arXiv",
    primaryClass = "gr-qc",
    doi = "10.1103/PhysRevD.97.104037",
    journal = "Phys. Rev. D",
    volume = "97",
    number = "10",
    pages = "104037",
    year = "2018"
}

@article{Belgacem:2018lbp,
    author = "Belgacem, Enis and Dirian, Yves and Foffa, Stefano and Maggiore, Michele",
    title = "{Modified gravitational-wave propagation and standard sirens}",
    eprint = "1805.08731",
    archivePrefix = "arXiv",
    primaryClass = "gr-qc",
    doi = "10.1103/PhysRevD.98.023510",
    journal = "Phys. Rev. D",
    volume = "98",
    number = "2",
    pages = "023510",
    year = "2018"
}

@article{Nishizawa:2019rra,
    author = "Nishizawa, Atsushi and Arai, Shun",
    title = "{Generalized framework for testing gravity with gravitational-wave propagation. III. Future prospect}",
    eprint = "1901.08249",
    archivePrefix = "arXiv",
    primaryClass = "gr-qc",
    doi = "10.1103/PhysRevD.99.104038",
    journal = "Phys. Rev. D",
    volume = "99",
    number = "10",
    pages = "104038",
    year = "2019"
}

@article{LIGOScientific:2016aoc,
    author = "Abbott, B. P. and others",
    collaboration = "LIGO Scientific, Virgo",
    title = "{Observation of Gravitational Waves from a Binary Black Hole Merger}",
    eprint = "1602.03837",
    archivePrefix = "arXiv",
    primaryClass = "gr-qc",
    reportNumber = "LIGO-P150914",
    doi = "10.1103/PhysRevLett.116.061102",
    journal = "Phys. Rev. Lett.",
    volume = "116",
    number = "6",
    pages = "061102",
    year = "2016"
}

@article{Bailes:2021tot,
    author = "Bailes, M. and others",
    title = "{Gravitational-wave physics and astronomy in the 2020s and 2030s}",
    doi = "10.1038/s42254-021-00303-8",
    journal = "Nature Rev. Phys.",
    volume = "3",
    number = "5",
    pages = "344--366",
    year = "2021"
}

@article{Cai:2017cbj,
    author = "Cai, Rong-Gen and Cao, Zhoujian and Guo, Zong-Kuan and Wang, Shao-Jiang and Yang, Tao",
    title = "{The Gravitational-Wave Physics}",
    eprint = "1703.00187",
    archivePrefix = "arXiv",
    primaryClass = "gr-qc",
    doi = "10.1093/nsr/nwx029",
    journal = "Natl. Sci. Rev.",
    volume = "4",
    number = "5",
    pages = "687--706",
    year = "2017"
}

@article{Regimbau:2011rp,
    author = "Regimbau, Tania",
    title = "{The astrophysical gravitational wave stochastic background}",
    eprint = "1101.2762",
    archivePrefix = "arXiv",
    primaryClass = "astro-ph.CO",
    doi = "10.1088/1674-4527/11/4/001",
    journal = "Res. Astron. Astrophys.",
    volume = "11",
    pages = "369--390",
    year = "2011"
}

@article{Christensen:2018iqi,
    author = "Christensen, Nelson",
    title = "{Stochastic Gravitational Wave Backgrounds}",
    eprint = "1811.08797",
    archivePrefix = "arXiv",
    primaryClass = "gr-qc",
    doi = "10.1088/1361-6633/aae6b5",
    journal = "Rept. Prog. Phys.",
    volume = "82",
    number = "1",
    pages = "016903",
    year = "2019"
}

@article{Renzini:2022alw,
    author = "Renzini, Arianna I. and Goncharov, Boris and Jenkins, Alexander C. and Meyers, Pat M.",
    title = "{Stochastic Gravitational-Wave Backgrounds: Current Detection Efforts and Future Prospects}",
    eprint = "2202.00178",
    archivePrefix = "arXiv",
    primaryClass = "gr-qc",
    doi = "10.3390/galaxies10010034",
    journal = "Galaxies",
    volume = "10",
    number = "1",
    pages = "34",
    year = "2022"
}

@inproceedings{Allen:1996vm,
    author = "Allen, Bruce",
    title = "{The Stochastic gravity wave background: Sources and detection}",
    booktitle = "{Les Houches School of Physics: Astrophysical Sources of Gravitational Radiation}",
    eprint = "gr-qc/9604033",
    archivePrefix = "arXiv",
    reportNumber = "WISC-MILW-96-TH-22",
    pages = "373--417",
    month = "4",
    year = "1996"
}

@article{Allen:1997ad,
    author = "Allen, Bruce and Romano, Joseph D.",
    title = "{Detecting a stochastic background of gravitational radiation: Signal processing strategies and sensitivities}",
    eprint = "gr-qc/9710117",
    archivePrefix = "arXiv",
    reportNumber = "WISC-MILW-97-TH-14",
    doi = "10.1103/PhysRevD.59.102001",
    journal = "Phys. Rev. D",
    volume = "59",
    pages = "102001",
    year = "1999"
}

@article{LIGOScientific:2025bgj,
    author = "Abac, A. G. and others",
    collaboration = "LIGO Scientific, VIRGO, KAGRA",
    title = "{Upper Limits on the Isotropic Gravitational-Wave Background from the first part of LIGO, Virgo, and KAGRA's fourth Observing Run}",
    eprint = "2508.20721",
    archivePrefix = "arXiv",
    primaryClass = "gr-qc",
    reportNumber = "LIGO-P2500349",
    month = "8",
    year = "2025"
}

@article{LIGOScientific:2025bkz,
    author = "Abac, A. G. and others",
    collaboration = "LIGO Scientific, VIRGO, KAGRA",
    title = "{Directional Search for Persistent Gravitational Waves: Results from the First Part of LIGO-Virgo-KAGRA's Fourth Observing Run}",
    eprint = "2510.17487",
    archivePrefix = "arXiv",
    primaryClass = "gr-qc",
    reportNumber = "LIGO-P250038",
    month = "10",
    year = "2025"
}

@article{NANOGrav:2023gor,
    author = "Agazie, Gabriella and others",
    collaboration = "NANOGrav",
    title = "{The NANOGrav 15 yr Data Set: Evidence for a Gravitational-wave Background}",
    eprint = "2306.16213",
    archivePrefix = "arXiv",
    primaryClass = "astro-ph.HE",
    doi = "10.3847/2041-8213/acdac6",
    journal = "Astrophys. J. Lett.",
    volume = "951",
    number = "1",
    pages = "L8",
    year = "2023"
}

@article{Agazie:2026tui,
    author = "Agazie, Gabriella and others",
    title = "{The NANOGrav 15 yr Data Set: Piecewise Power-Law Reconstruction of the Gravitational-Wave Background}",
    eprint = "2601.09481",
    archivePrefix = "arXiv",
    primaryClass = "astro-ph.HE",
    month = "1",
    year = "2026"
}

@article{NANOGrav:2025gqp,
    author = "Agarwal, Nikita and others",
    collaboration = "NANOGrav",
    title = "{The NANOGrav 15 yr Dataset: Targeted Searches for Supermassive Black Hole Binaries}",
    eprint = "2508.16534",
    archivePrefix = "arXiv",
    primaryClass = "astro-ph.HE",
    doi = "10.3847/2041-8213/ae3719",
    journal = "Astrophys. J. Lett.",
    volume = "998",
    number = "1",
    pages = "L11",
    year = "2026"
}

@article{ET:2025xjr,
    author = "Abac, Adrian and others",
    collaboration = "ET",
    title = "{The Science of the Einstein Telescope}",
    eprint = "2503.12263",
    archivePrefix = "arXiv",
    primaryClass = "gr-qc",
    reportNumber = "ET-0036C-25",
    doi = "10.1088/1475-7516/2026/03/081",
    journal = "JCAP",
    volume = "03",
    pages = "081",
    year = "2026"
}

@article{Evans:2021gyd,
    author = "Evans, Matthew and others",
    title = "{A Horizon Study for Cosmic Explorer: Science, Observatories, and Community}",
    eprint = "2109.09882",
    archivePrefix = "arXiv",
    primaryClass = "astro-ph.IM",
    reportNumber = "CE-P2100003-v7, Cosmic Explorer technical report CE-P2100003-v6",
    month = "9",
    year = "2021"
}

@article{LISA:2022yao,
    author = "Seoane, Pau Amaro and others",
    collaboration = "LISA",
    title = "{Astrophysics with the Laser Interferometer Space Antenna}",
    eprint = "2203.06016",
    archivePrefix = "arXiv",
    primaryClass = "gr-qc",
    doi = "10.1007/s41114-022-00041-y",
    journal = "Living Rev. Rel.",
    volume = "26",
    number = "1",
    pages = "2",
    year = "2023"
}

@article{Berti:2015itd,
    author = "Berti, Emanuele and others",
    title = "{Testing General Relativity with Present and Future Astrophysical Observations}",
    eprint = "1501.07274",
    archivePrefix = "arXiv",
    primaryClass = "gr-qc",
    doi = "10.1088/0264-9381/32/24/243001",
    journal = "Class. Quant. Grav.",
    volume = "32",
    pages = "243001",
    year = "2015"
}

@article{Ishak:2018his,
    author = "Ishak, Mustapha",
    title = "{Testing General Relativity in Cosmology}",
    eprint = "1806.10122",
    archivePrefix = "arXiv",
    primaryClass = "astro-ph.CO",
    doi = "10.1007/s41114-018-0017-4",
    journal = "Living Rev. Rel.",
    volume = "22",
    number = "1",
    pages = "1",
    year = "2019"
}

@article{Clifton:2011jh,
    author = "Clifton, Timothy and Ferreira, Pedro G. and Padilla, Antonio and Skordis, Constantinos",
    title = "{Modified Gravity and Cosmology}",
    eprint = "1106.2476",
    archivePrefix = "arXiv",
    primaryClass = "astro-ph.CO",
    doi = "10.1016/j.physrep.2012.01.001",
    journal = "Phys. Rept.",
    volume = "513",
    pages = "1--189",
    year = "2012"
}

@article{LIGOScientific:2026qni,
    author = "Abac, A. G. and others",
    collaboration = "LIGO Scientific, VIRGO, KAGRA",
    title = "{GWTC-4.0: Tests of General Relativity. I. Overview and General Tests}",
    eprint = "2603.19019",
    archivePrefix = "arXiv",
    primaryClass = "gr-qc",
    reportNumber = "LIGO-P2500065",
    month = "3",
    year = "2026"
}

@article{LIGOScientific:2026fcf,
    author = "Abac, A. G. and others",
    collaboration = "LIGO Scientific, VIRGO, KAGRA",
    title = "{GWTC-4.0: Tests of General Relativity. II. Parameterized Tests}",
    eprint = "2603.19020",
    archivePrefix = "arXiv",
    primaryClass = "gr-qc",
    reportNumber = "LIGO-P2500066",
    month = "3",
    year = "2026"
}

@article{LIGOScientific:2026wpt,
    author = "Abac, A. G. and others",
    collaboration = "LIGO Scientific, VIRGO, KAGRA",
    title = "{GWTC-4.0: Tests of General Relativity. III. Tests of the Remnants}",
    eprint = "2603.19021",
    archivePrefix = "arXiv",
    primaryClass = "gr-qc",
    reportNumber = "LIGO-P2500067",
    month = "3",
    year = "2026"
}

@article{Nunes:2020rmr,
    author = "Nunes, Rafael C.",
    title = "{Searching for modified gravity in the astrophysical gravitational wave background: Application to ground-based interferometers}",
    eprint = "2007.07750",
    archivePrefix = "arXiv",
    primaryClass = "gr-qc",
    doi = "10.1103/PhysRevD.102.024071",
    journal = "Phys. Rev. D",
    volume = "102",
    number = "2",
    pages = "024071",
    year = "2020"
}

@article{Ezquiaga:2021ayr,
    author = "Ezquiaga, Jose Mar{\'\i}a",
    title = "{Hearing gravity from the cosmos: GWTC-2 probes general relativity at cosmological scales}",
    eprint = "2104.05139",
    archivePrefix = "arXiv",
    primaryClass = "astro-ph.CO",
    doi = "10.1016/j.physletb.2021.136665",
    journal = "Phys. Lett. B",
    volume = "822",
    pages = "136665",
    year = "2021"
}

@article{Phinney:2001,
    author = "Phinney, E. S.",
    title = "{A Practical Theorem on Gravitational Wave Backgrounds}",
    eprint = "astro-ph/0108028",
    archivePrefix = "arXiv",
    year = "2001"
}

@article{Renzini2024popstock,
  author  = {Renzini, Arianna I. and Golomb, Jacob},
  title   = {Projections of the uncertainty on the compact binary population
             background using popstock},
  journal = {Astron. Astrophys.},
  volume  = {691},
  pages   = {A238},
  year    = {2024},
  eprint  = {2407.03742},
  archivePrefix = {arXiv},
}

@article{Giarda2025,
doi = {10.1088/1361-6382/ae07a0},
url = {https://doi.org/10.1088/1361-6382/ae07a0},
year = {2025},
month = {sep},
publisher = {IOP Publishing},
volume = {42},
number = {19},
pages = {195015},
author = {Giarda, Giovanni and Renzini, Arianna I and Pacilio, Costantino and Gerosa, Davide},
title = {Accelerated inference of binary black-hole populations from the stochastic gravitational-wave background},
journal = {Classical and Quantum Gravity},
}

@article{Khan2016,
  author  = {Khan, Sebastian and Husa, Sascha and Hannam, Mark and
             Ohme, Frank and P{\"u}rrer, Michael and
             Jim{\'e}nez Forteza, Xisco and Boh{\'e}, Alejandro},
  title   = {{IMRPhenomD}: Phenomenological gravitational waveforms
             for binary black holes},
  journal = {Phys. Rev. D},
  volume  = {93},
  pages   = {044007},
  year    = {2016},
}

@article{Abbott2023pop,
  author  = {Abbott, R. and others},
  collaboration = {LIGO Scientific, Virgo, KAGRA},
  title   = {Population of merging compact binaries inferred using
             gravitational waves through {GWTC-3}},
  journal = {Phys. Rev. X},
  volume  = {13},
  pages   = {011048},
  year    = {2023},
}

@article{Madau2014,
  author  = {Madau, Piero and Dickinson, Mark},
  title   = {Cosmic Star-Formation History},
  journal = {Annu. Rev. Astron. Astrophys.},
  volume  = {52},
  pages   = {415--486},
  year    = {2014},
}

@article{Reitze2019,
  author  = {Reitze, David and others},
  title   = {Cosmic Explorer: The U.S.\ Contribution to
             Gravitational-Wave Astronomy beyond {LIGO}},
  journal = {Bull. Am. Astron. Soc.},
  volume  = {51},
  pages   = {35},
  year    = {2019},
}

@article{Meacher2015,
  author  = {Meacher, D. and others},
  title   = {Mock data and science challenge for detecting an astrophysical
             stochastic gravitational-wave background with {Advanced LIGO}
             and {Advanced Virgo}},
  journal = {Phys. Rev. D},
  volume  = {92},
  pages   = {063002},
  year    = {2015},
}

@article{Planck2015,
  author  = {Ade, P. A. R. and others},
  collaboration = {Planck},
  title   = {{Planck} 2015 results. {XIII}.\ Cosmological parameters},
  journal = {Astron. Astrophys.},
  volume  = {594},
  pages   = {A13},
  year    = {2016},
}

@article{Talbot2019,
  author  = {Talbot, Colm and Smith, Rory and Thrane, Eric and Poole, Greg B.},
  title   = {Parallelized inference for gravitational-wave astronomy},
  journal = {Phys. Rev. D},
  volume  = {100},
  pages   = {043030},
  year    = {2019},
}

@article{Renzini2023pygwb,
  author  = {Renzini, Arianna I. and Romero-Rodr{\'i}guez, Alba and Talbot, Colm and Lalleman, Max and Kandhasamy, Shivaraj and others},
  title   = {{pygwb}: {A} {Python}-based Library for Gravitational-wave Background Searches},
  journal = {The Astrophysical Journal},
  year    = {2023},
  volume  = {952},
  number  = {1},
  pages   = {25},
  doi     = {10.3847/1538-4357/acd775},
  eprint  = {2303.15696},
  archivePrefix = {arXiv},
  primaryClass  = {gr-qc}
}

@article{Meacher:2015iua,
    author = "Meacher, Duncan and Coughlin, Michael and Morris, Sean and Regimbau, Tania and Christensen, Nelson and Kandhasamy, Shivaraj and Mandic, Vuk and Romano, Joseph D. and Thrane, Eric",
    title = "{Mock data and science challenge for detecting an astrophysical stochastic gravitational-wave background with Advanced LIGO and Advanced Virgo}",
    eprint = "1506.06744",
    archivePrefix = "arXiv",
    primaryClass = "astro-ph.HE",
    doi = "10.1103/PhysRevD.92.063002",
    journal = "Phys. Rev. D",
    volume = "92",
    number = "6",
    pages = "063002",
    year = "2015"
}

@article{Regimbau:2022mdu,
    author = "Regimbau, Tania",
    title = "{The Quest for the Astrophysical Gravitational-Wave Background with Terrestrial Detectors}",
    doi = "10.3390/sym14020270",
    journal = "Symmetry",
    volume = "14",
    number = "2",
    pages = "270",
    year = "2022"
}

@article{Ashton_2019,
   title={Bilby: A User-friendly Bayesian Inference Library for
					Gravitational-wave Astronomy},
   volume={241},
   ISSN={1538-4365},
   url={http://dx.doi.org/10.3847/1538-4365/ab06fc},
   DOI={10.3847/1538-4365/ab06fc},
   number={2},
   journal={The Astrophysical Journal Supplement Series},
   publisher={American Astronomical Society},
   author={Ashton, Gregory and Hübner, Moritz and Lasky, Paul D. and Talbot, Colm and Ackley, Kendall and Biscoveanu, Sylvia and Chu, Qi and Divakarla, Atul and Easter, Paul J. and Goncharov, Boris and Vivanco, Francisco Hernandez and Harms, Jan and Lower, Marcus E. and Meadors, Grant D. and Melchor, Denyz and Payne, Ethan and Pitkin, Matthew D. and Powell, Jade and Sarin, Nikhil and Smith, Rory J. E. and Thrane, Eric},
   year={2019},
   month=Apr, pages={27} }

@article{Mancarella:2021ecn,
    author = "Mancarella, Michele and Genoud-Prachex, Edwin and Maggiore, Michele",
    title = "{Cosmology and modified gravitational wave propagation from binary black hole population models}",
    eprint = "2112.05728",
    archivePrefix = "arXiv",
    primaryClass = "gr-qc",
    doi = "10.1103/PhysRevD.105.064030",
    journal = "Phys. Rev. D",
    volume = "105",
    number = "6",
    pages = "064030",
    year = "2022"
}

@article{LIGOScientific:2025jau,
    author = "Abac, A. G. and others",
    collaboration = "LIGO Scientific, VIRGO, KAGRA",
    title = "{GWTC-4.0: Constraints on the Cosmic Expansion Rate and Modified Gravitational-wave Propagation}",
    eprint = "2509.04348",
    archivePrefix = "arXiv",
    primaryClass = "astro-ph.CO",
    reportNumber = "LIGO-P2400152",
    month = "9",
    year = "2025"
}

@misc{LIGOScientific:2026uyd,
      title={GWTC-5.0: Constraints on the Cosmic Expansion Rate and Modified Gravitational-wave Propagation}, 
      author={The LIGO Scientific Collaboration and the Virgo Collaboration and the KAGRA Collaboration},
      year={2026},
      eprint={2605.27227},
      archivePrefix={arXiv},
      primaryClass={astro-ph.CO},
      url={https://arxiv.org/abs/2605.27227}, 
}

@article{Lagos:2019kds,
    author = "Lagos, Macarena and Fishbach, Maya and Landry, Philippe and Holz, Daniel E.",
    title = "{Standard sirens with a running Planck mass}",
    eprint = "1901.03321",
    archivePrefix = "arXiv",
    primaryClass = "astro-ph.CO",
    doi = "10.1103/PhysRevD.99.083504",
    journal = "Phys. Rev. D",
    volume = "99",
    number = "8",
    pages = "083504",
    year = "2019"
}

@article{Alonso:2016suf,
    author = "Alonso, David and Bellini, Emilio and Ferreira, Pedro G. and Zumalac{\'a}rregui, Miguel",
    title = "{Observational future of cosmological scalar-tensor theories}",
    eprint = "1610.09290",
    archivePrefix = "arXiv",
    primaryClass = "astro-ph.CO",
    reportNumber = "NORDITA-2016-114",
    doi = "10.1103/PhysRevD.95.063502",
    journal = "Phys. Rev. D",
    volume = "95",
    number = "6",
    pages = "063502",
    year = "2017"
}

@article{Bellini:2015xja,
    author = "Bellini, Emilio and Cuesta, Antonio J. and Jimenez, Raul and Verde, Licia",
    title = "{Constraints on deviations from {\ensuremath{\Lambda}}CDM within Horndeski gravity}",
    eprint = "1509.07816",
    archivePrefix = "arXiv",
    primaryClass = "astro-ph.CO",
    doi = "10.1088/1475-7516/2016/06/E01",
    journal = "JCAP",
    volume = "02",
    pages = "053",
    year = "2016",
    note = "[Erratum: JCAP 06, E01 (2016)]"
}

@article{Saffer:2020xsw,
    author = "Saffer, Alexander and Yagi, Kent",
    title = "{Parameter Estimation for Tests of General Relativity with the Astrophysical Stochastic Gravitational Wave Background}",
    eprint = "2003.11128",
    archivePrefix = "arXiv",
    primaryClass = "gr-qc",
    doi = "10.1103/PhysRevD.102.024001",
    journal = "Phys. Rev. D",
    volume = "102",
    number = "2",
    pages = "024001",
    year = "2020"
}

\end{document}